\documentclass[sigconf,10pt]{acmart}
\usepackage{amsmath}
\usepackage{amsfonts}
\usepackage{amssymb}
\usepackage{bm}
\usepackage{comment}
\usepackage[ruled,vlined]{algorithm2e}
\renewcommand{\CommentSty}[1]{\textnormal{#1}}
\DontPrintSemicolon
\SetKwComment{tcp}{$\triangleright$ }{}
\SetVlineSkip{0cm}
\usepackage{enumitem}

\renewcommand{\algorithmcfname}{Alg.}%

\usepackage{colortbl}
\usepackage{multicol}

\usepackage{subcaption}

\usepackage{balance}

\def\BibTeX{{\rm B\kern-.05em{\sc i\kern-.025em b}\kern-.08emT\kern-.1667em\lower.7ex\hbox{E}\kern-.125emX}}

\copyrightyear{2020} 
\acmYear{2020} 
\setcopyright{acmcopyright}\acmConference[SIGMOD'20]{Proceedings of the 2020 ACM SIGMOD International Conference on Management of Data}{June 14--19, 2020}{Portland, OR, USA}
\acmBooktitle{Proceedings of the 2020 ACM SIGMOD International Conference on Management of Data (SIGMOD'20), June 14--19, 2020, Portland, OR, USA}
\acmPrice{15.00}
\acmDOI{10.1145/3318464.3389761}
\acmISBN{978-1-4503-6735-6/20/06}

\begin{document}

\title{SLIM: Scalable Linkage of Mobility Data}

\author{Fuat Bas{\i}k}
\authornote{Work done prior to author's employment with Amazon. This work is included in the author's PhD thesis.}
\email{fbbasik@amazon.co.uk}
\orcid{0000-0001-6190-0203}
\affiliation{%
  \institution{Amazon Web Services}
}

\author{Hakan Ferhatosmano\u{g}lu}
\email{hakan.f@warwick.ac.uk}
\orcid{0000-0002-5181-4712}
\affiliation{%
  \institution{University of Warwick}
}

\author{Bu\u{g}ra Gedik}
\email{bgedik@cs.bilkent.edu.tr}
\orcid{0000-0002-0622-1141}
\affiliation{%
  \institution{Bilkent University}
}

\begin{abstract}
We present a scalable solution to link entities across mobility datasets using their spatio-temporal information. This is a fundamental problem in many applications such as linking user identities for security, understanding privacy limitations of location based services, or producing a unified dataset from multiple sources for urban planning. 
Such integrated datasets are also essential for service providers to optimise their services and improve business intelligence. In this paper, we first propose a mobility based representation and similarity computation for entities. An efficient matching process is then developed to identify the final linked pairs, with an automated mechanism to decide when to stop the linkage. We scale the process with a locality-sensitive hashing (LSH) based approach that significantly reduces candidate pairs for matching. To realize the effectiveness and efficiency of our techniques in practice, we introduce an algorithm called SLIM. In the experimental evaluation, SLIM outperforms the two existing state-of-the-art approaches in terms of precision and recall. Moreover, the LSH-based approach brings two to four orders of magnitude speedup.
\end{abstract}


\maketitle

\section{Introduction}
\noindent With the proliferation of smart phones integrated with positioning systems and the increasing penetration of Internet-of-Things (IoT), mobility data has become widely available~\cite{ref:4point}. A vast variety of mobile services and applications have a location-based context and produce spatio-temporal records. For example, most payments produce records containing the time of the payment and the location of the store. 
Location sharing services such as Swarm
help people share their whereabouts. These records contain information about both the entities and the environment they were produced in. Availability of such data supports smart services in areas including healthcare~\cite{ref:pentlandHealth},
computational social sciences~\cite{ref:pentland2009}, and location-based marketing~\cite{ref:mstutorial}. 

There are many studies that analyze location based data for social good, such as reducing traffic congestion, lowering air pollution levels, analyzing the spread of contagious diseases, and modelling of the pandemic virus spreading process~\cite{ref:crowdsensing,ref:airline, ref:virus}. Most of the studies focus on a single dataset, which provides only a partial and biased state, failing to capture the complete patterns of mobility. To produce a comprehensive view of mobility, one needs to integrate multiple datasets, potentially from disparate sources. Such integration enables knowledge extraction that cannot be obtained from a single data source, and benefits a wide range of applications and machine learning tasks~\cite{ref:combTaxi, ref:trajectoryDataMining}. Examples include discovering regional poverty by jointly using mobile and satellite data
where accurate demographic information is not available~\cite{ref:poverty}, and measuring urban social diversity by jointly modeling the attributes from Twitter and Foursquare~\cite{ref:musolesiwww16}. 

Spatio-temporal linkage is necessary to avoid over- or under-estimation of population densities using multiple sources of data, e.g., signals from wifi based positioning and mobile applications. It is also essential in user identification for security purposes, and understanding privacy consequences of anonymized mobility data~\cite{ref:gm2}. An outcome of work such as ours is to help developing privacy advisor tools where location based activities are assessed in terms of their user identity linkage likelihood.

Identifying the matching entities across mobility datasets is a non-trivial task, since some datasets are anonymized due to privacy or business concerns, and hence unique identifiers are often missing. Since spatio-temporal information is the only feature guaranteed to exist in all mobility datasets, our approach 
decreases the number of features the linkage depends on, and can be used for other applications. This helps avoid the use of personally identifying information (PII)~\cite{ref:www} and additional sensitive data, and simplifies the procedures to share data for social good and research purposes without having to explicitly expose PII. It enables use and collection of minimal data, which is a requirement in major privacy and data protection regulations, such as GDPR. As such, in this paper, we present a scalable solution to link entities across mobility datasets, using only their spatio-temporal information. There are several challenges associated with such linkage across mobility datasets.

First, a summary representation for the spatio-temporal records and an associated similarity measure should be developed.
The similarity score should capture the closeness, in time and location, of the records. Yet, unlike the traditional measures, it should not penalize the absence of records in one dataset for a particular time window, when the other dataset has records for the same time window. This is because different services are typically not used synchronously.

Second, once similarity scores are assigned to entity pairs, an efficient matching process needs to identify the final linked pairs. A challenging problem is to automate the decision to stop the linkage to avoid false positive links. In a real-world setting, it is unlikely to have the entities from one dataset as a subset of the other, which is a commonly made assumption.
This is an important but so far overseen issue in the literature~\cite{ref:automatic, ref:gm, ref:trajectoryRecovery}.

Third, efficiency and scalability of the linkage process are essential, given the scale and dynamic nature of location datasets. Even a static comparison of each pair of entities would require quadratic number of similarity computations over millions of entities.  Avoiding exhaustive search and focusing on a small amount of pairs that are likely to be matching can scale the linkage to a larger number of entities for real-life dynamic cases.

In this work, we present a scalable linkage solution, 
\emph{SLIM}, for efficiently finding the matching entities across mobility datasets, relying only on the spatio-temporal information.

Given an entity, we introduce a \emph{mobility history} representation, by distributing the recorded locations over time-location bins (Sec.~\ref{sec:preliminaries}) and defining a novel similarity score for histories, based on a scaled aggregation of the proximities of their matching bins (Sec.~\ref{sec:sim-score}). 
The proposed similarity definition provides several important properties. It awards the matching of close time-location bins and incorporates the frequency of the bins in the award amount. It normalizes the similarity scores based on the size of the histories in terms of the number of time-location bins. Moreover, while it does not penalize the score when one entity has activity in a particular time window but the other does not, it does penalize the existence of cross-dataset activities that are close in time but distant in space (aka \emph{alibi}s~\cite{ref:stlink, ref:phd}). This is an essential property that supports mobility linkage.

The similarity scores are used as weights to construct a bipartite graph which is used for maximum sum matching. The matched entities are linked via an automated linkage thresholding (Sec.~\ref{sec:matching}). We first fit a mixture model, e.g., Gaussian Mixture Model (GMM), with two components over the distribution of the edge weights selected by the maximum sum bipartite matching. One of these components aims to model the true positive links and the other one is for false positive links. We then formulate the expected precision, recall, and $F1$-score for a given threshold, based on the fitted model. This is used to automatically select the threshold that provides the maximum $F1$-Score and use it to filter the results to produce the linkage.

To address the scalability challenge, we develop a \emph{Locality Sensitive Hashing} (LSH)~\cite{ref:lsh} based approach using the concept of \emph{dominating grid cell} (Sec.~\ref{sec:lsh}). Dominating grid cells contain most records of the owner entity during a given time period. We construct a list of dominating cells to act as signatures and then apply a banding technique, by dividing the signatures into $b$ bands consisting of $r$ rows, where each band is hashed to a large number of buckets. The goal is enable signatures such that entities with similarity higher than a threshold $t$ are hashed to the same bucket at least once. Our experimental evaluation (Sec.~\ref{sec:eval}) shows that this LSH based approach brings two to four orders of magnitude speedup to linkage with a slight reduction in the recall.

In summary, this work makes the following contributions:%
\smallskip\\
$\bullet\,$\emph{Model}. We devise a hierarchical summary representation for mobility records of entities along with a novel method to compute their similarity. The similarity score we introduce
does not only capture the closeness in time and location, but also tolerates temporal asynchrony and penalizes the \emph{alibi} records, which are essential contributions for mobility linkage.
\smallskip\\
$\bullet\,$\emph{Algorithm}. We develop the SLIM algorithm for linking entities across mobility datasets. It relies on maximum bipartite matching over a graph formed using the similarity scores. We provide an automated mechanism to detect an appropriate threshold to stop the linkage; a crucial step for avoiding false positives.
\smallskip\\
$\bullet\,$\emph{Scalability}. To scale linkage to large datasets, we develop an \emph{LSH} based approach which brings a significant speedup. To our knowledge, this is the first application of LSH in the context of mobility linkage.
\smallskip\\
$\bullet\,$\emph{Evaluation}. We perform extensive experimental evaluation using real-world datasets, compare SLIM with two existing approaches (STLink~\cite{ref:stlink}, GM~\cite{ref:gm}), and show superior performance in terms of accuracy and scalability.
\section{Preliminaries}
\label{sec:preliminaries}
\noindent 
In this section, we first present the notation used throughout the paper. We then formally define the linkage problem, and describe how mobility histories are represented.

\subsection{Notation}
\noindent
\textbf{\emph{Location Datasets}}. A location dataset is defined as a collection of usage records from a location-based service. We use $\mathcal{I}$ and $\mathcal{E}$ to denote the two location datasets from the two services, across which the linkage will be performed. 
\smallskip

\noindent
\textbf{\emph{Entities}}. Entities are real-world systems or users, whose usage records are stored in the location datasets. They are represented in the datasets with their ids. An id uniquely identifies an entity within a dataset. However, since ids are anonymized, they can be different across different datasets and cannot be used for linkage. The set of all entities of a location dataset is represented as $U_\mathcal{E}$, where the subscript represents the dataset. Throughout this work, we use $u \in U_\mathcal{E}$ and $v \in U_\mathcal{I}$ to represent two entities from the two datasets.%
\smallskip
\\
\noindent
\textbf{\emph{Records}}. Records are generated as a result of entities interacting with the location-based service. Each record is a triple $\{u,l,t\}$, where for a record $r\in\mathcal{E}$, $r.u$ represents the id of the entity associated with the record. Similarly, $r.l$ and $r.t$ represent the location and timestamp of the record, respectively. We assume that the record locations are given as points. Our approach can be extended to datasets that contain record locations as regions, by copying a record into multiple cells within the mobility histories using weights.

\subsection{Problem Definition}
With these definitions at hand, we can define the problem as follows. Given two location datasets, $\mathcal{I}$ and $\mathcal{E}$, the problem is to find a one-to-one mapping from a subset of the entities in the first set to a subset of the entities in the second set. This can be more symmetrically represented as a function that takes a pair of entities, one from the first dataset and a second from the other, and returns a Boolean result that either indicates a positive linkage (true) or no-linkage (false), with the additional constraint that an entity cannot be linked to more than one entity from the other dataset. A positive linkage indicates that the relevant entities from the two datasets refer to the same entity in real-life. 

More formally, we are looking for a linkage function $\mathcal{M}: U_\mathcal{E} \times U_\mathcal{I} \rightarrow \{0, 1\}$, with the following constraint:
\begin{align*}
\exists\, u, v & \text{ s.t. } \mathcal{M}(u, v) = 1\\
 &\Rightarrow \forall u^\prime \neq u, v^\prime \neq v,\, \mathcal{M}(u^\prime, v) = \mathcal{M}(u, v^\prime) = 0
\end{align*}

Note that the size of the overlap between sets of entities from different services may not be known in advance. Even when it is known, finding all positive linkages is often not possible, as some of the entities may not have enough records to establish linkage. 

\subsection{Mobility Histories}
\begin{figure}[t]
\centering
\includegraphics[width=0.75\linewidth]{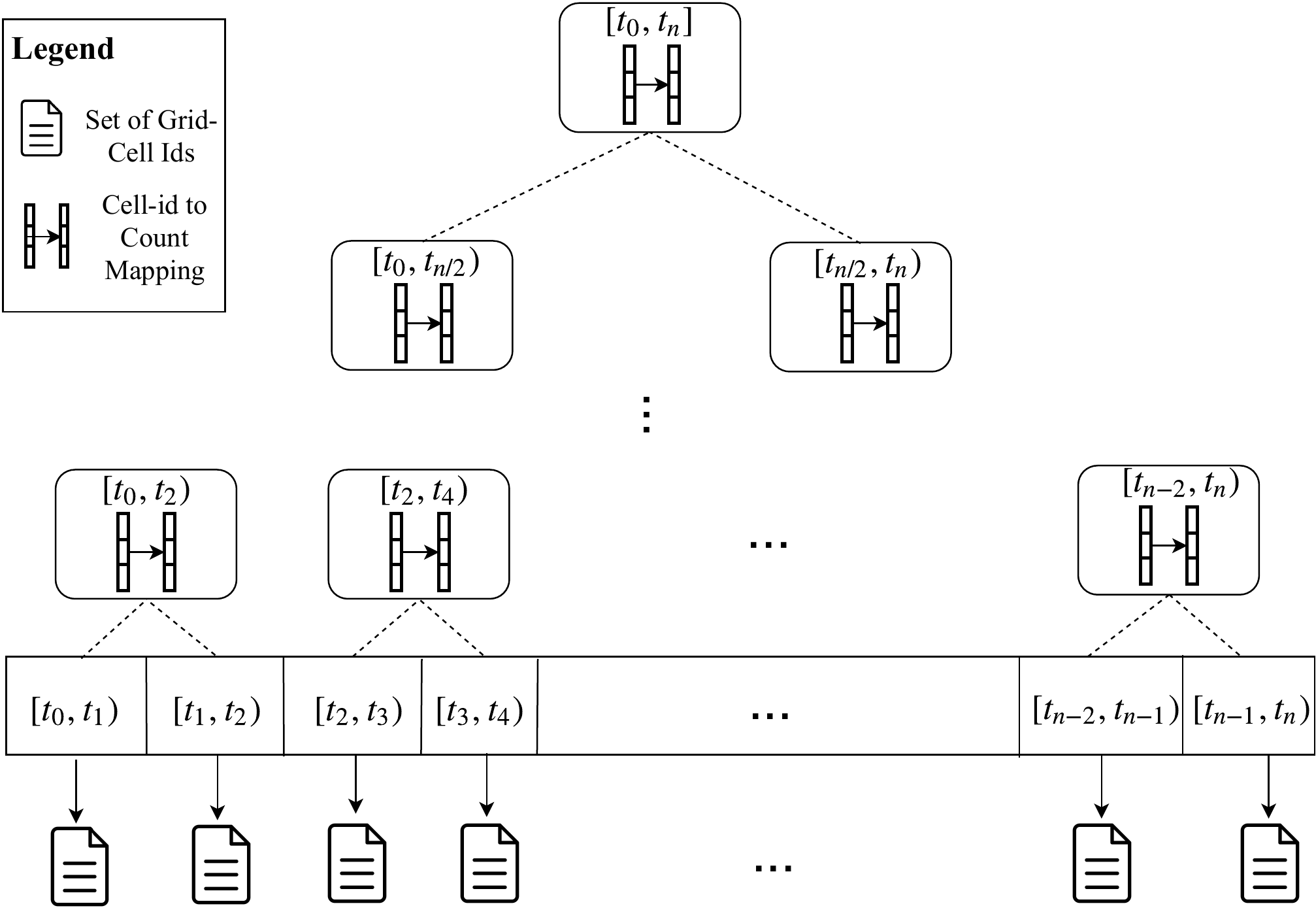}
\caption{Mobility history representation}
\label{fig:profile}
\end{figure}
\noindent 
We organize the records in a location dataset into \emph{mobility histories}. Given an entity, its mobility history consists of an aggregated collection of its records from the dataset. Representation of the mobility history should be generic enough to capture the differences and spatio-temporal heterogeneity of the datasets. To address this need, we propose a hierarchical representation for mobility histories, which distributes the records over time-location bins.

In the temporal domain, the data is split into time windows which are hierarchically organized as a tree to enable efficient computation of aggregate statistics. The leaves of this tree hold a set of spatial cell ids. A cell id is present at a leaf node if the entity has at least one record whose spatial location is in that cell and the record timestamp is in the temporal range of the window. Each non-leaf node keeps the occurrence counts of the cell ids in its sub-tree. The space complexity of this tree is similar to a segment tree, $\mathcal{O}(|\mathcal{E}| + |\mathcal{I}|)$. As we will detail in Sec.~\ref{sec:lsh}, the information kept in the non-leaf nodes is used for scalable linkage.  The cells are part of spatial partitioning of locations. For this, we use S2\footnote{\url{http://s2geometry.io/}}, which divides the Earth's surface into 31 levels of hierarchical cells, where, at the smallest granularity, the leaf cells cover $1 cm^2$. 

We deliberately form the mobility history tree via hierarchical temporal partitioning, and not via hierarchical spatial partitioning. This is because spatial partitioning is not effective in detecting negative linkage (alibi~\cite{ref:stlink}). Given two locations in the same temporal window, if it is not possible for an entity to move from one of these locations to the other within the width of the window, then these two records are considered as a negative link, i.e., alibi. %Since our similarity score calculation takes alibis into consideration, organizing the mobility histories by hierarchically partitioning in the temporal domain is more effective. 
To calculate the similarity score of an entity pair, we compare their records to those that are in close temporal proximity. Record pairs that are close in both time and space are awarded, whereas record pairs that are close in time but distant in space are penalized. Hence, we favor fast retrieval of records based on temporal information over based on spatial information.

Both the temporal window size used for the leaf nodes and the S2 level (spatial granularity) used for the cells are configurable. As detailed later, we auto-tune the spatial granularity for a given temporal window size using the trade-off between accuracy and performance of the linkage.

\subsection{Overview of the Linkage Process}
\noindent Figure~\ref{fig:profile} shows an example mobility history of an entity. Each leaf keeps a set of locations, represented with cell ids. Each non-leaf node keeps the information on how many times a cell id has occurred at the leaf level in its sub-tree.

The linkage is performed in three steps. First, a Locality-Sensitive Hashing based filtering step reduces the number of entity pairs that needs to be considered for linkage. The second step computes the pairwise similarity scores of entities from the two datasets. The similarity scores are computed based on a formula that determines how similar the mobility histories are. The computed similarities are used as weights to construct a bipartite graph of the entities from the two datasets. The final step is to apply a maximum sum bipartite matching, where the matched entities are considered as linked. Once the matches are found, they are output as linkage only if their scores are above the \emph{stop similarity threshold}, which is automatically identified.
\section{Mobility History Linkage}
\label{sec:linkage}
\noindent In this section we describe the computation of similarity scores of mobility history pairs and the maximum weight bipartite matching for linkage. Since the set of entities in the two location datasets are not necessarily identical and the amount of intersection between them is not known in advance, we also discuss how to find an appropriate score threshold to limit the number of entities linked.

\subsection{Mobility History Similarity Score}
\label{sec:sim-score}
\noindent  The similarity score of a pair of mobility histories should capture closeness in time and location. This score should not require a consistent matching of time-location bins across two histories. This is because the mobility histories are not synchronous, i.e., a record is not necessarily present for the same timestamp in both of the datasets.  Therefore, the similarity score needs to aggregate the proximities of the time-location bins, different from the traditional techniques where similarity measures are defined over records, like Minkowski distance or Jaccard similarity.  We define a number of desired properties for the similarity score:

1) Award the matching of close time-location bins. While the similarity score contribution of exactly matching time-location bins is obvious, one should also consider bins that are from the same temporal window but are from different yet spatially close cells. Such time-location bins are deemed close and they should contribute to the similarity score relative to their closeness.

2) Tolerate temporal asynchrony, that is, do not penalize the similarity when one entity has records in a particular time window but the other does not. This case is common when the location datasets are obtained from asynchronous services. While spatio-temporally close usage should contribute to the similarity, lack of it for a particular time window should not penalize the score.

3) Penalize alibi time-location bins. Time-location bins that are from the same time window but whose cells are so distant in space that it is not possible for an entity to move between these cells within the time window are considered as alibi bins. They are counter evidences in terms of linking the entities and penalized in the similarity score. 

4) Award infrequent cells in matching time-location bins. While summing up the proximities of the matched time-location bins, the uniqueness of the cells should be awarded as they are stronger indicators of similarity than cells that are seen frequently.   

5) Normalize the similarity score contributions based on the size of the mobility histories in terms of the number of time-location bins. If not handled properly, the skew in the number of time-location bins would result in the mobility histories with too many bins to dominate the similarity scores over the shorter ones. 

\subsubsection{Proximity of Time-Location Bins}
\noindent One of the desired properties of the time-location bin proximity was tolerating temporal asynchrony. 
Hence, we consider only the temporally close time-location bins in our proximity computation. 

Let $\mathcal{T}$ be a binary function that takes the value of $1$ if the two given time-location bins are associated with the same temporal window and $0$ otherwise. Let $e = \{w, c\}$ and $i = \{w, c\}$ be two time-location bins where $e.w$ and $i.w$ are two temporal windows and $e.c$  and $i.c$ are two spatial grid cells. We define $\mathcal{T}$ formally as $\mathcal{T}(e,i) = \bm{[}e.w = i.w\bm{]}$.

We define the spatial proximity of a pair of time-location bins as an inverse function of the geographical distance of their locations. However, we use an upper bound on the geographical distance to capture the concept of alibis. This upper bound, referred to as the \emph{runaway distance}, is defined as the maximum distance an entity can travel within the given temporal window.  It is represented as $\mathcal{R}$ and calculated by multiplying the width of the temporal window with the maximum speed, $\alpha$, at which an entity can travel. Let $w$ be a temporal window and $|w|$ be the width of it, then $\mathcal{R} = |w| \cdot \alpha$. In practice, the value of $\alpha$ could be defined using a dataset-specific speed limit.
We define the proximity function $\mathcal{P}$ formally as follows:
\begin{equation}
 \label{eq:proximity}
 \mathcal{P}(e,i) =  \mathcal{T} (e,i) \cdot \log_{2}{(2 - min(d(e.c, i.c)/\mathcal{R}, 2))}
\end{equation}
where $d$ is a function that calculates the minimum geographical distance between two grid cells.

When a given pair of time-location bins are not from the same temporal window, the outcome of this function becomes $0$. When they are from the same temporal window and the same spatial cell, the outcome becomes $1$ --- the highest value it can take. As the distance increases up to the runaway $R$, the value goes down to $0$ with an increasing slope. If the distance is more than $R$, the outcome becomes negative, reaching $-\infty$ as the distance reaches two times the runaway distance. This is a simple yet effective technique to capture the alibi record pairs. In a real-world setting, the location data might be inaccurate. Therefore, while the decrease to negative values is steep, it is still a continuous function that allows a small number of alibi record pairs whose distance from each other is slightly larger than the runaway distance. 

The proximity function $\mathcal{P}$ is designed so that our similarity score satisfies the first three of the required properties we have outlined earlier, namely: awarding the matching of close time-location bins, tolerating temporal asynchrony, and penalizing the alibis.

\subsubsection{Aggregation of Proximities}
\noindent The similarity computation is performed over the time-location bins at the leaves of the histories. For entity $u\in U_{\mathcal{E}}$ (and $v\in U_{\mathcal{I}}$), we represent the set of time-locations bins as $H_u$ (and $H_v$), where $e\in H_u = \{c, w\}$ (and  $i\in H_v = \{c, w\}$) represents a time-location bin with the time window $w$ and the grid cell $c$. 

Given a pair of mobility histories, the computation starts with constructing pairs of time-location bins whose proximity will be computed and included in the aggregation.

The design of mobility history trees blocks the records by distributing them over time-location bins. This is similar to sorted neighborhood based blocking of traditional record linkage algorithms~\cite{ref:cristenSurvey}. Normally, pairs of time-location bins would have been constructed using Cartesian product over events from the same temporal window. However, this would be over-counting, as a time-location bin will end up participating in multiple pairs. Therefore, we first introduce a pairing function $\mathcal{N}(u, v)$, which computes the set of time-location bin pairs to be included in the aggregation for a pair $(u, v)$ of mobility histories. Later, in the experimental evaluation, we show that this pairing function $N$ is more accurate compared to all pairs matching.

We restrict ourselves to time-location bin pairs from corresponding time windows, as this guarantees that the pairs satisfy the temporal proximity, $\mathcal{T}$. As such, we have $\mathcal{N}(u, v) = \bigcup_w \mathcal{N}_w(u, v) $. Given a time window $w$, we compute $\mathcal{N}_w(u,v)$ by first selecting the time-location bin pair $(e, i)$ with $e.w = i.w = w$ that has the smallest geographical distance $d(e.c, i.c)$.  
Once this pair is added into $\mathcal{N}_w(u,v)$, pairs containing any one of the selected time-location bins, that is $e$ or $i$, are removed from the remaining set of candidate pairs. We repeat these two operations until there are no time-location bins left in the smaller mobility history.  

Once the set of time-location bin pairs to include in the aggregation are found, we define the similarity score function of two mobility histories for the entities $u\in U_\mathcal{E}$ and $v\in U_\mathcal{I}$, as follows:
\begin{eqnarray}
\mathcal{S}(u,v) =& \sum\limits_{\{e,i\} \in \mathcal{N}(u, v)} \mathcal{P}(e,i)\, \frac{\min{(idf(e, \mathcal{E}), idf(i, \mathcal{I}))}}{\mathcal{L}(u,\mathcal{E})  \mathcal{L}(v,\mathcal{I})}\nonumber\\
\mbox{where:}&\mathcal{L}(u,\mathcal{E}) = (1-b) + b \frac{|H_u|}{\sum_{u'\in U_\mathcal{E}} |H_{u'}|/|U_\mathcal{E}|}\label{eq:sim}
\end{eqnarray}

The similarity function, $\mathcal{S}$, has three main components. The first component is the proximity $\mathcal{P}$, as defined in Eq.~\ref{eq:proximity}. For the time-location bin pairs identified by the pairing function $\mathcal{N}$, we sum up the proximities. The other two components deal with the scaling of the proximity value, in order to $i$) normalize the differences in the mobility history sizes in terms of the bin counts, and $ii$) award infrequent cells in the matching. These two properties implement the last two desired properties from Sec.~\ref{sec:sim-score}.

Histories with a higher number of time-location bins, compared to other histories from the same location dataset, would be more likely to be similar with the histories from the opposite dataset. We introduce a normalization function $\mathcal{L}$, for both histories, which makes the contribution of each bin pair to the similarity score inversely proportional with the relative sizes of the histories. The relative size of a mobility history is defined as the ratio of the number of time-location bins it contains over the average number of time-location bins from the same dataset. 

In order to tune the impact of the mobility history size in terms of time-location bins, we add a parameter $b$, which takes a value between 0 and 1. At one extreme, $b=0$, the denominator becomes $1$, i.e., the history lengths are ignored. At the other end, $b=1$, the denominator becomes the product of the relative history sizes. This is inspired by BM25~\cite{ref:bm25} --- a ranking function used in document retrieval, which avoids bias towards long documents. 

The final component of the similarity function is the \emph{idf} multiplier. This component is analogous to inverse document frequency in information retrieval. It awards uniqueness of a pair of time-location bins. If an entity is in an infrequent time-location bin, i.e., the number of other entities from the same dataset in the same time-location bin is low, the contribution of this bin to the similarity score should be higher. Likewise, if a time-location bin is highly common among entities of one dataset, the contribution should be lower. As the frequency of time-location bins might differ across datasets, we take the \textit{idf} score that makes the lowest contribution. $idf$ of a time-location bin equals to the logarithm of the ratio of the number of mobility histories to the number of mobility histories that contain the given time location bin. Formally, given a time location bin, $e\in H_u$, for the entity $u\in U_\mathcal{E}$, we calculate $idf(e, \mathcal{E})$ as follows:
\begin{equation}
idf(e, \mathcal{E}) = \log{(|U_{\mathcal{E}}|/|\{u\in U_\mathcal{E}\mid e \in H_u\}|)}
\end{equation}

\begin{algorithm}[!t]
	\begin{small}
 	\KwData{ $\mathcal{E}, \mathcal{I}$: Location datasets to be linked}
 	\KwResult{$\mathcal{L}$: Linked pairs of entities}
 	$\{\mathcal{H}_\mathcal{E}\ , \mathcal{H}_\mathcal{I}\} \leftarrow CreateHistories(\mathcal{E}, \mathcal{I})$ \tcp*{Histories from  datasets}
    $E \leftarrow \{\} \, , \, V \leftarrow \{\}$ \tcp*{Initialize edges and vertices}
    %$$ \tcp*{Initialize vertices}
 	\For(\tcp*[f]{For each history in first history set}){$H_u \in \mathcal{\mathcal{H}_\mathcal{E}}$}{
    	$\mathcal{W}_{H_u} \leftarrow H_u.getAllWindows()$ \tcp*{Get all windows of $H_u$}
        \For(\tcp*[f] {For a candidate history}){$H_v \in LSHFilterPairs(u)$}{
            $S \leftarrow 0$ \tcp*{Initialize similarity score}
			$\mathcal{W}_{H_v} \leftarrow H_v.getAllWindows()$ \tcp*{Get all windows of $H_v$}
            \For(\tcp*[f]{For a common window}){$w \in W_{H_u} \cap W_{H_v}$}{
                $N \leftarrow \mathcal{N}_w(u, v)$ \tcp*{Mutually nearest pairs}
                $S \leftarrow S + \mathcal{S}(N)$ \tcp*{Update similarity (see Eq.~\ref{eq:sim}\footnotemark)}
                $N' \leftarrow \mathcal{N'}_w(u,v)$ \tcp*{Mutually furthest pairs}
                \For(\tcp*[f]{For each mutually furthest pair}){$(e, i) \in N'$}{
                    $D \leftarrow \mathcal{S}(\{(e, i)\})$ \tcp*{Delta similarity}
                    \If(\tcp*[f]{If an alibi is detected}){$D < 0$}{
            	        $S \leftarrow S + D$ \tcp*{Update similarity}
                    }
                }
            }
            \If(\tcp*[f]{If score is positive}){$S > 0$}{
                $V \leftarrow V \cup \{u, v\}$ \tcp*{Add to vertex set}
                $E \leftarrow E \cup \{(u,v; S)\}$  \tcp*{Add to weighted edge set}
        	}
        }
	}
	%$\mathcal{L} \leftarrow  \tcp*{Find linked entities}
	\textbf{return} $\mathcal{L} \leftarrow LinkPairs(E,V)$ \tcp*{Return linked entity pairs}
	\caption{SLIM: Scalable Linkage of Mobility Histories}
  	\label{algo:slim}	
	\end{small} 	
\end{algorithm}
\footnotetext{We overload the notation such that $\mathcal{S}$ takes the bin pairs as input.}

Algorithm~\ref{algo:slim}, SLIM, shows the linkage of the mobility histories using our similarity score.
SLIM starts by creating the mobility histories from the location datasets. It then finds the mutually nearest neighbor (MNN) pairs for each corresponding time window ($\mathcal{N}$). The similarity score is computed by aggregating the weighted proximities for these pairs, as it was outlined in Eq.~\ref{eq:sim}. 

An additional step that uses mutually furthest neighbors (MFN) further improves the effectiveness of alibi detection. We illustrate this with an example. Given a temporal window, let $e_1$ be an entity with a single time-location bin $b_1$ and $e_2$ be an entity with two such bins, $b_2$ and $b_3$. Assume that the distance between $b_1$ and $b_2$ is $d$ units and the distance between $b_1$ and $b_3$ is $d+r$ units, where $d<r$ and $r$ is the runaway distance. While MNN would return the pair with distance $d$ (missing the alibi), MFN would help us capture the alibi time-location bin pair. This is shown in the algorithm with the inner-most for loop, where $\mathcal{N}'$ (MFN) function acts similar to $\mathcal{N}$ (MNN), but it chooses the pairs with the furthest distance. To avoid double counting, we \emph{only} consider these pairs if they are alibis.

Once the similarity scores are computed for the mobility history pairs, they are used to construct a weighted bipartite graph. If the score is negative, no edges are added to the graph. Next, we describe how to perform maximum sum bipartite matching and decide a stop point for the linkage.

\subsection{Final Linkage}
\label{sec:matching}
\noindent The SLIM algorithm computes a weighted bipartite graph $G(E,V)$, where $V = U_\mathcal{E} \cup U_\mathcal{I}$ and $E$ is the set of edges between them. This bipartite  graph is used to find a maximum weighted matching where the two ends of the selected edges are considered linked. To avoid ambiguity in the results, the selected edges should not share a vertex. Since the matching is performed on a bipartite graph, there is no edge between two entities from the same dataset.

Finding this matching is a combinatorial optimization problem known as the \emph{assignment problem}, with many optimal and approximate solutions~\cite{ref:hungarian, ref:approximations}. We adapt a simple greedy heuristic, which links the pair with the highest similarity at each step. Maximum weighted matching algorithms are traditionally designed to find a full matching. In other words, all entities from the smaller set of entities will be linked to an entity from the larger set. However, in a real-world linkage setting, it is unlikely to have the entities from one dataset to be a subset of the other. Frequently, it is not even possible to know the intersection amount in advance. This is an important but so far an overseen issue in the related literature~\cite{ref:automatic, ref:gm, ref:trajectoryRecovery}. For this, we design a mechanism to decide an appropriate threshold score to stop the linkage, to avoid false positive links when the sets of entities from two location datasets are not identical. 

After performing a full matching over the bipartite graph, there are two sets of selected edges. The first set is the true positive links, which contains the selected edges whose entities at the two ends refer to the same entity in real life. The second set is the false positive links, which contains the rest of the selected edges. The purpose of a threshold for stopping the linkage is to increase the precision, by ruling out some of the selected edges from the result set. Ideally, this should be done without harming the recall, by extracting edges only from the false positive set. However, this is a challenging task when ground truth does not exist.

A good similarity score should group true positive links and false positive links in two clusters that are distinguishable from each other. With the assumption that our similarity score satisfies these properties, to determine the stop threshold, we first fit a 1-dimensional Gaussian Mixture Model (GMM) with two components over the distribution of the selected edge weights~\cite{ref:gmm}. One can adopt more sophisticated models for this purpose. We assume that the two components ($m_1$ and $m_2$) have independent Gaussian distribution of weights and the component with the larger mean ($m_2$) models the true positive links (the other modelling the false positive links). Assume that there is already a similarity score threshold $s$. The cumulative distribution functions of the components $m_1$ and $m_2$ are used to compute: $i$) the area under the curve of $m_1$ and to the right of $y=s$ line, which gives the rate of false positives, and $ii$) the area under the curve of $m_2$ and and to the right of $y=s$ line, which gives the rate of true positives. 

\begin{figure}[t]
\centering
\includegraphics[width=0.65\linewidth]{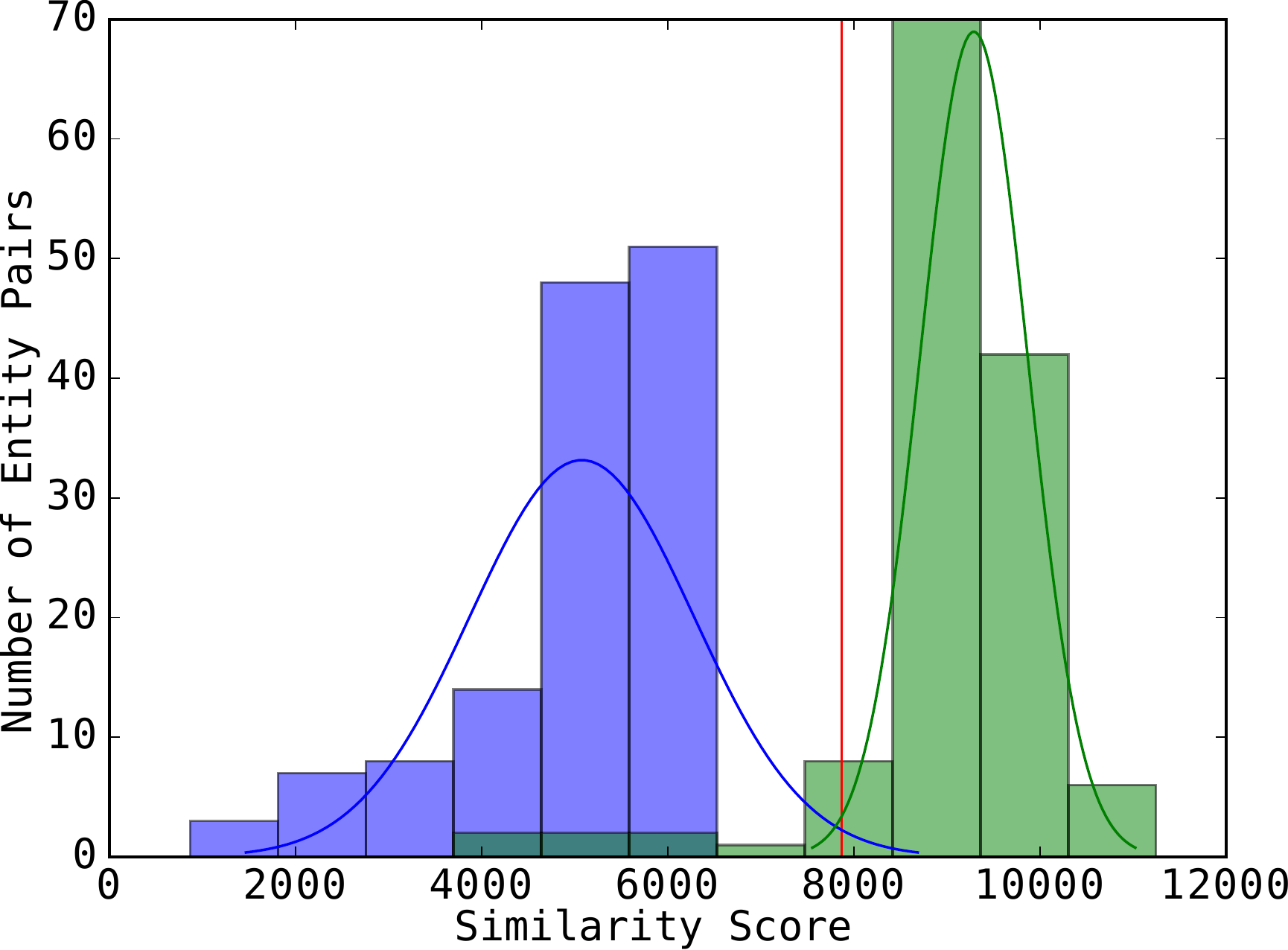}
\caption{Sample GMM fit for similarity scores}
\label{fig:gmm}
\end{figure}

Using precision ($P(s)$), and recall ($R(s)$), one could calculate a combined F1-score as $F1(s) = 2 P(s) R(s)/(P(s) + R(s))$. Note that all these scores are dependent on the score threshold $s$. Let $c_1$ ad $c_2$ denote the weights of the GMM components $m_1$ and $m_2$, respectively. We have $R(s) = c_2 (1 - F_{m_2}(s))$ and $P(s) = R(s) / (R(s) + c_1 (1 - F_{m_1}(s)))$, where $F$ represents the cumulative distribution function. Finally, we have $s^* = argmin_s F1(s)$ as the linkage stop score threshold to use.

We only include the links whose edges are higher than the threshold in the result. Figure~\ref{fig:gmm} shows an example of GMM fitting on sample similarity scores. The $x$-axis shows the scores and the $y$-axis shows the number of links in a particular score bin. The two lines show the components of the GMM. The green bars in the histogram show the number of true positive links and the blue bars show the number of false positive links. %(obtained from the ground truth and shown for illustrative purposes). 
Vertical red line is the detected linkage stop similarity score threshold value. Note that, this is not a supervised approach and the ground truth is used only for illustrative purposes.

\subsection{Performance Tuning}\label{sec:tune}
\noindent Existing work identify width of the temporal window, and the spatial level of detail either using previously labeled data~\cite{ref:gm} or use preset values identified via human intuition~\cite{ref:www}. Here, we take a step forward to automatically tune the spatial level for a given temporal window width, in the absence of previously labeled data. 

The trade-off when deciding the spatial level is between accuracy and performance. When the spatial domain is coarsely partitioned, the record locations become indistinguishable. On the other hand, increasing the spatial detail, hence creating finer grained partitions of space, increases the size of the mobility histories. Linking  larger histories takes more processing time.
Yet, increasing the spatial detail does not always improve the accuracy of the linkage. Our observations based on experiments with multiple datasets show that, after a certain level of detail, increasing the spatial detail neither improves nor worsens the accuracy of the linkage. 

To find out the best spatial grid level that balances the aforementioned trade-off, we perform a test on distinguishing entities from the same dataset. When the level of detail is too low, similarity scores of all pairs would be close to each other. In this case, the similarity score of entity pairs $u$ and $v$ will be close to the similarity score of $u$ and $u$ (self-similarity). Using higher details of spatial information would decrease this ratio, indicating an entity is more similar to itself than any other entity. Making use of this observation, we first select a subset of entities from the dataset and form a set of pair of entities by crossing them with the rest of the entities. Next, for changing spatial level of detail, we compute the average of the aforementioned ratio (pair similarity over self-similarity) for all pairs. Once we have the average values for each spatial detail, we perform a best trade-off point detection algorithm (aka. elbow point detection) as implemented in~\cite{ref:knee}. Repeating this procedure independently for two datasets to be linked, we use the higher elbow point as the spatial detail level of the linkage. In our experimental evaluation we show that this technique is able to detect most accurate spatial detail level that does not add overhead in the performance, for a given temporal window.
\section{Scalable Linkage}
\label{sec:lsh}
\noindent Unlike the traditional approaches, sorting mobility histories or finding a blocking key to split entities into multiple blocks~\cite{ref:cristenSurvey} is not possible in mobility history linkage. Due to the number of events each history contains, the cost of calculating similarities of mobility histories and comparison based blocking techniques are computationally impractical~\cite{ref:blockingSurvey}. 
Therefore, we utilise \emph{Locality-Sensitive Hashing} (LSH)~[15] for efficient identification of pairs that are likely to be similar. A challenging task here is to design the LSH and the corresponding similarity for the addressed linkage problem.

For asynchronous and sparse datasets, representing the mobility histories as sets of \emph{k-shingles} of records and expecting identical bands to apply min-hashing~\cite{ref:momd} would be overly strict. On the other hand, for two entities to link it is expected that most of their records are generated in the same spatial grid cell. These \emph{dominating grid cells}~\cite{ref:stlink} contain most records of the owner entity and are determined by simply picking the cell with the highest count of the entity's records. While one dominating grid cell could be found using the entire dataset, it is also possible to specify a start and end time to form a query, and find the dominating cell for a particular time window defined by this query. 

Given a mobility history, we construct a list of dominating grid cells to act as signatures. This is done by querying each history for non-overlapping time windows to find the corresponding dominating cells, and adding the resulting cells to the end of the signature. We make sure that the queries span the same time period with the data, and the order of the queries is the same across histories. If a history does not contain any record for a query time window, a unique placeholder is added to the signature to make sure signatures have the same structure. In other words, the dominating grid cells obtained from the same index of different signatures are results of the same query. When applying hashing, these placeholders are omitted.

Each mobility history is converted into a signature consisting of a sequence of dominating grid cells. The similarity $t$ across two signatures is defined as the number of matching dominating cells, divided by the signature size. The signature size is determined by the query window size, which is a multiple of the leaf-level window size and a parameter of our LSH procedure. The appropriate level of the mobility history tree is used to quickly locate the dominating cells. 

With the signatures at hand, we apply the banding technique, like in the case of document matching. The signatures are divided into $b$ bands consisting of $r$ rows, and each band is hashed to a large number of buckets. The goal here is to come up with a setting such that signatures with similarity higher than a threshold $t$ to be hashed to the same bucket at least once. Given two signatures of similarity $t$, the probability of these signatures having at least one identical band is $1-(1-t^r)^b$.  Regardless of the constants $b$ and $r$,  this function has an S-Curve shape and the threshold $t$ is the point where the rise is the steepest. Consequently, it is possible to approximate the threshold $t$ as $(\frac{1}{b})^{\frac{1}{r}}$~\cite{ref:momd}.

The number of bands, $b$, that reaches a particular threshold, $t$, can be calculated as follows. Let $s$ be the signature size and $t$ be the similarity threshold for becoming a candidate pair. Given $t=(1/b)^{1/r}$ and $r = s/b$, we have $t = (1/b)^{b/s}$. Solving for $b$ gives: $b = e^{\mathcal{W}(-s \ln{t})}$
where $\mathcal{W}$ is the \emph{Lambert W} function, which is the inverse of the function $f(x) = x\, e^x$~\cite{ref:LambertW}.
The parameters to the LSH procedure for mobility histories are: $i$) the similarity threshold $t$, $ii$) the query time window size for determining dominating cells, and $iii$) the spatial level of the dominating cells. 

\noindent\textbf{Illustrative Example.}
\begin{figure}[t]
\centering
\includegraphics[width=0.85\linewidth]{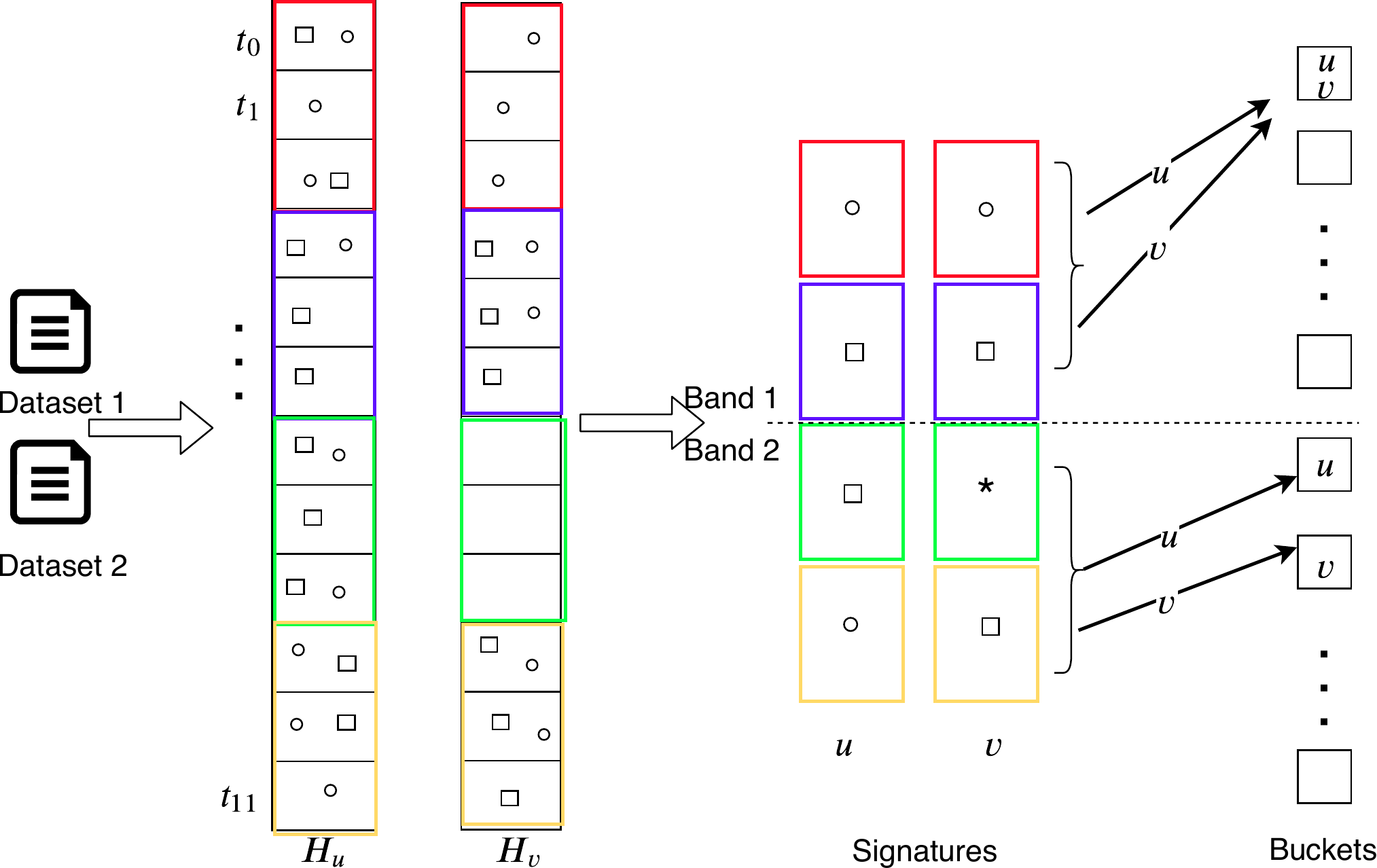}
\caption{LSH of mobility histories}
\label{fig:lsh}
\end{figure}
\noindent The collection of records is first converted into mobility histories (Figure~\ref{fig:profile}). Leaf levels of the histories for two entities, $u$ and $v$, are shown in Figure~\ref{fig:lsh} ($H_u$, $H_v$). In this representation, each history consists of $12$ time windows. 
For illustration, we assume that entities visit only two spatial cells, represented with \emph{square} and \emph{circle}.

To apply LSH, we first query the mobility histories four times to identify the dominating cells. Each query has a window size of three time units and is shown in different colors. The resulting signatures are of length $4$. For the first query (red rectangle), the entity $u$ visited the grid cell \emph{circle} $3$ times and it visited the cell \emph{square} $2$ times. Therefore, the dominating cell for entity $u$ during the first query time window is grid-cell \emph{circle}. This procedure is repeated for all queries and entities. The third index of the signature for entity $v$ has the mark $*$ because it has no records during the third query's time window (green rectangle). Once the signatures are ready, we apply the banding technique using two bands. For the first band, since the signatures are identical, the entities are hashed to the same bucket. 

Since the entities $u$ and $v$ are candidate pairs, we next compute their similarity score. In the first time window (the upper most cell), $u$ has visited two distinct cells, \emph{circle} and \emph{square}, and entity $v$ visited \emph{circle} only. The pairing function $\mathcal{N}$ takes these two sets of locations and computes the set of record pairs to be included in the aggregation. In this example, this function pairs \emph{circles}, as the distance between them are minimum (i.e., they are mutually nearest neighbors). The contribution of this pair to the similarity score is computed according to Eq~\ref{eq:sim} that takes proximity of the cells, their popularity and the length of histories into account. Last, the algorithm would check if there is an alibi pair of events in the given set. This time, we use  $\mathcal{N}'$ as the pairing function and it pairs \emph{square} and \emph{circle} grid cells as they are mutually furthest neighbors of these two sets of records. Depending on the geographical proximity of these two cells, this pair might have a negative contribution to the aggregation (alibi). This procedure will be repeated independently for each temporal window and the computed scores for each time window will be summed. Once the computation is finished, the resulting score is used to set the weight of the edge between the entities $u$ and $v$ in the final bipartite graph.

The last step is to perform the maximum weighted bipartite matching and to select a subset of the resulting edges. Once the edges and their weights are determined, we fit a 1-dimensional GMM with two components over the weights of the edges and determine the stop threshold to use for linkage. We report only the edges that have higher weight than this stop threshold as a link.
\section{Experimental Evaluation}
\label{sec:eval}
\noindent In this section, we give a detailed evaluation of the proposed solution, SLIM and compare it with state-of-the-art. We implemented SLIM using Java 1.8. All experiments were executed on a Linux server with 2 Intel Xeon E5520 2.27GHz CPUs and 64GB of RAM.

\subsection{Datasets}
\begin{figure*}[ht]
	\begin{subfigure}{0.24\linewidth}
		\includegraphics[width=\linewidth]{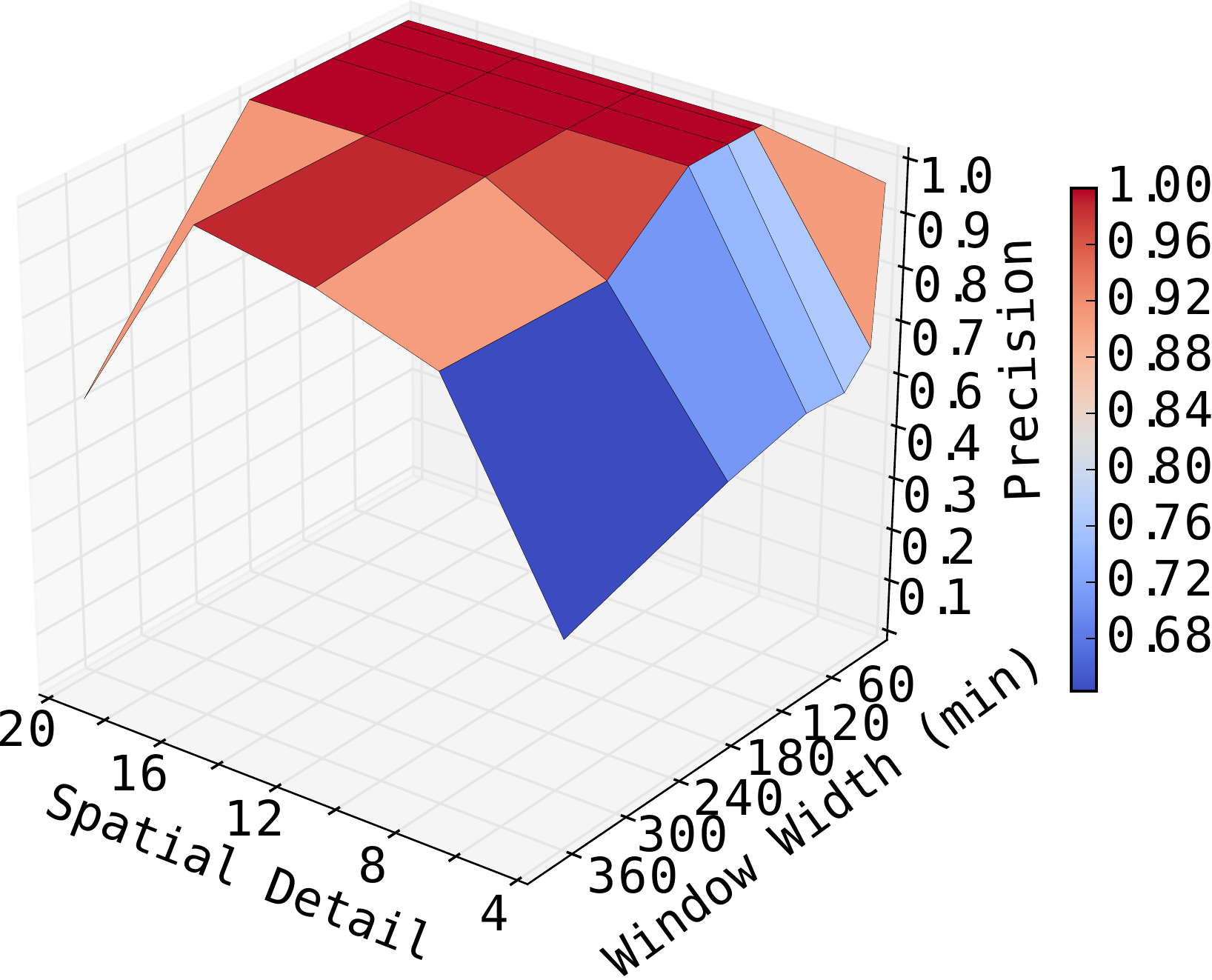}
 		\caption{Precision}
		\label{fig:1-1cabprecision}
	\end{subfigure}
	\hspace{0.0001\linewidth}
  	\begin{subfigure}{0.24\linewidth}
		\includegraphics[width=\linewidth]{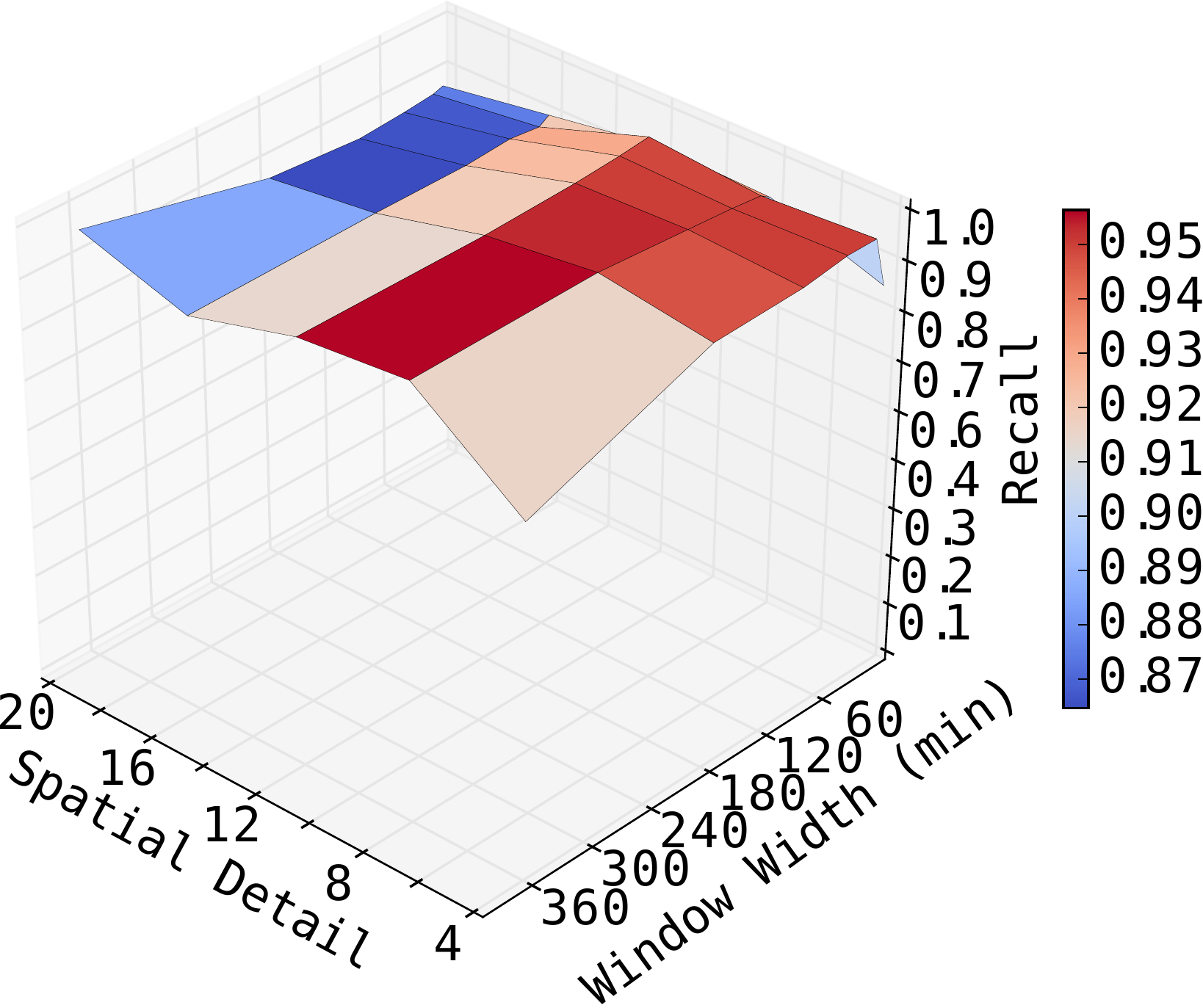}
 		\caption{Recall}
		\label{fig:1-1cabrecall}
	\end{subfigure}
  	\hspace{0.0001\linewidth}
	\begin{subfigure}{0.24\linewidth}
  		\includegraphics[width=\linewidth]{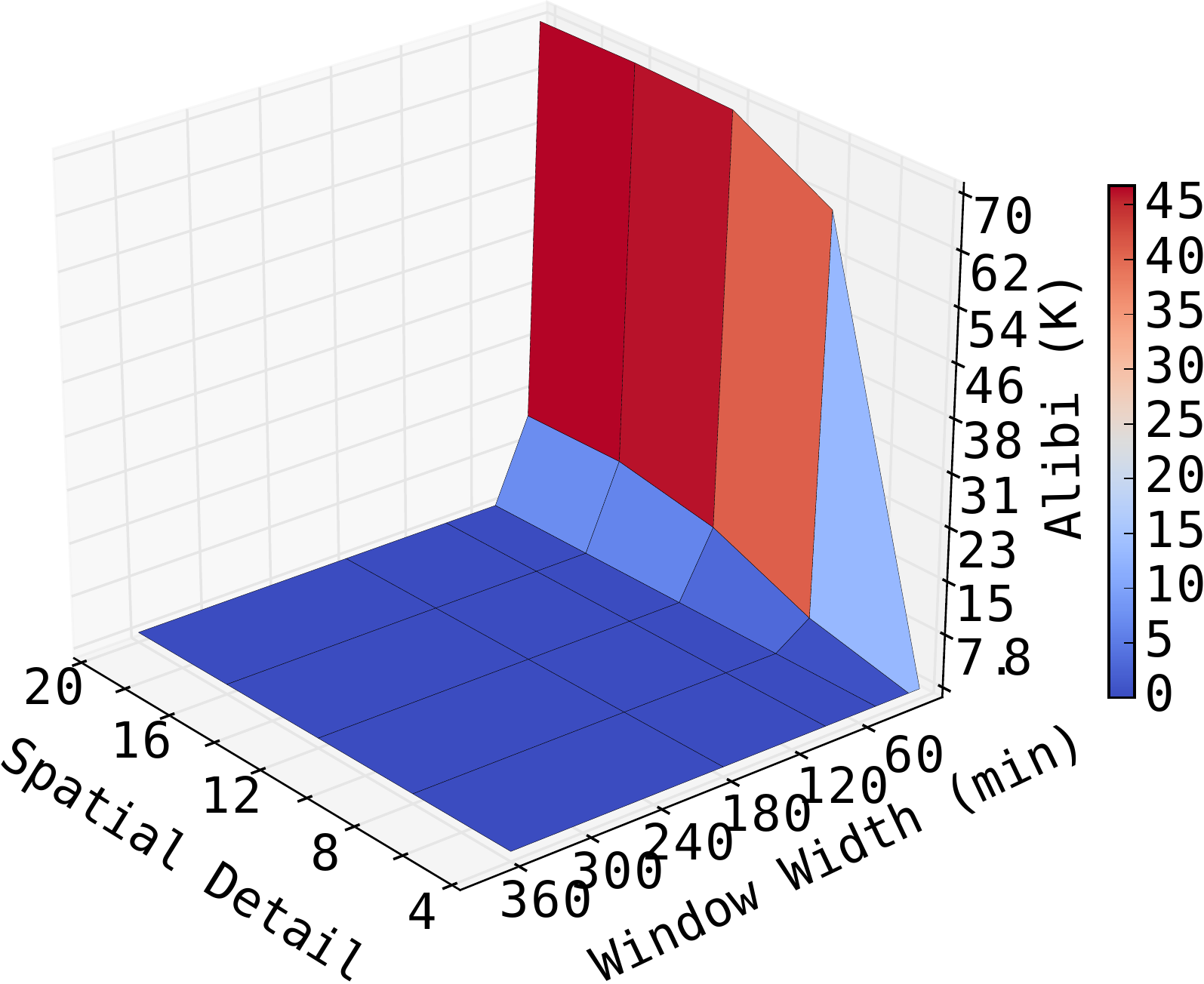}
  		\caption{\# of alibi pairs}
		\label{fig:1-1cabalibi}
	\end{subfigure}
	\hspace{0.0001\linewidth}
	\begin{subfigure}{0.24\linewidth}
  		\includegraphics[width=\linewidth]{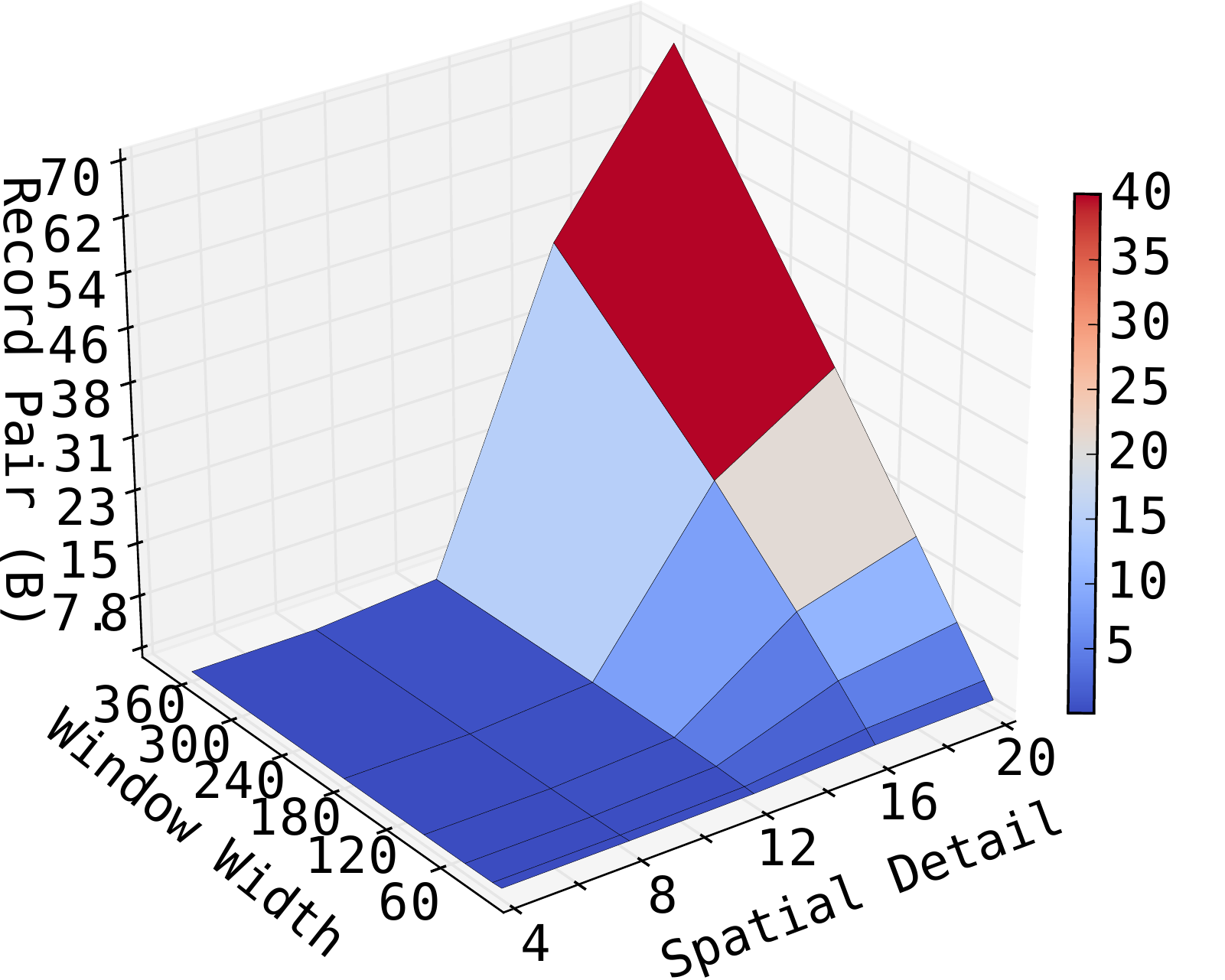}
  		\caption{\# of event comparisons}
		\label{fig:1-1cabevent}
	\end{subfigure}
	\caption{Effect of the spatio-temporal level -- Cab}
	\label{fig:1-1cab}
\end{figure*}
\begin{figure*}[ht]
	\begin{subfigure}{0.24\linewidth}
		\includegraphics[width=\linewidth]{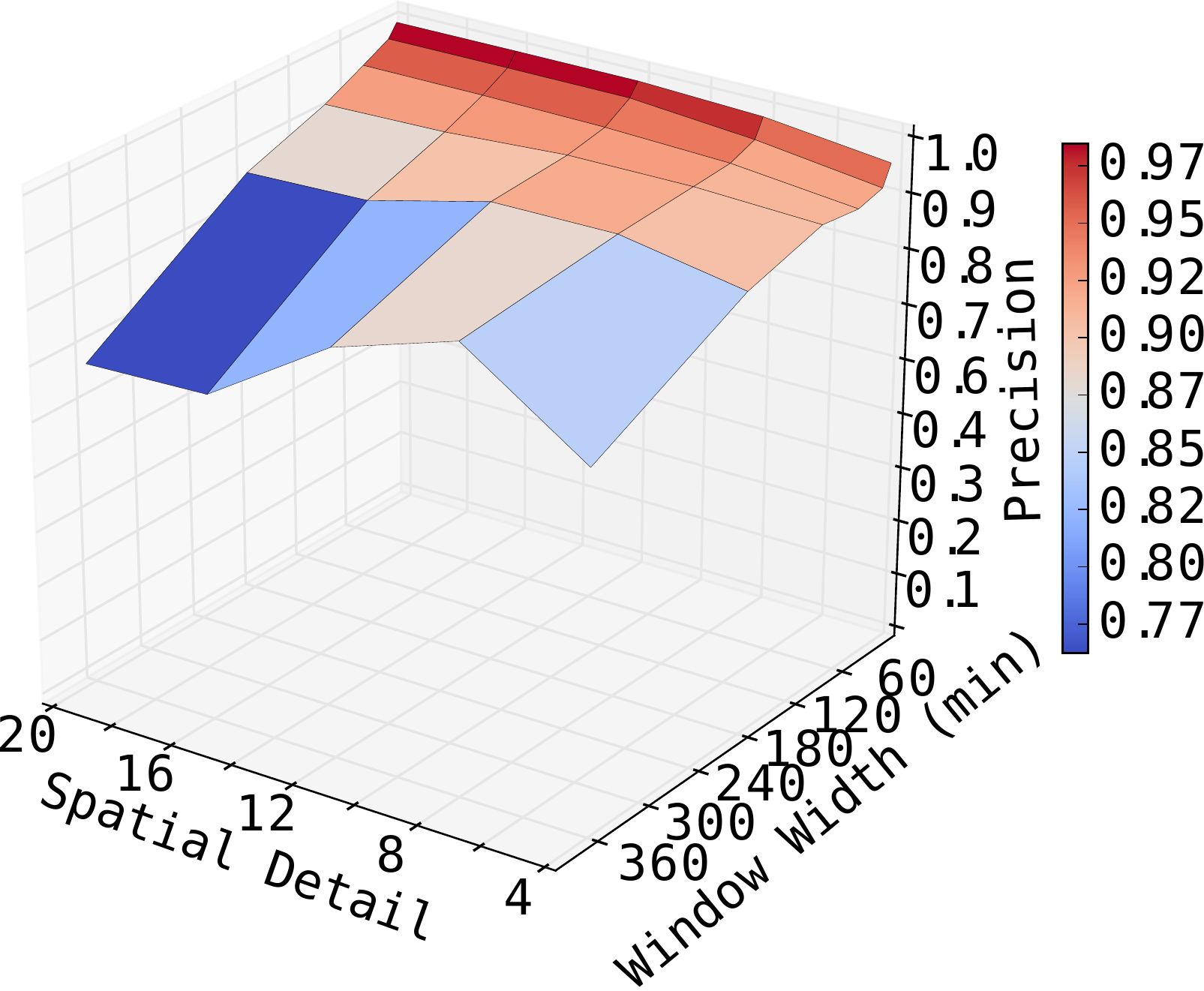}
 		\caption{Precision}
		\label{fig:1-1smprecision}
	\end{subfigure}
	\hspace{0.0001\linewidth}
  	\begin{subfigure}{0.24\linewidth}
		\includegraphics[width=\linewidth]{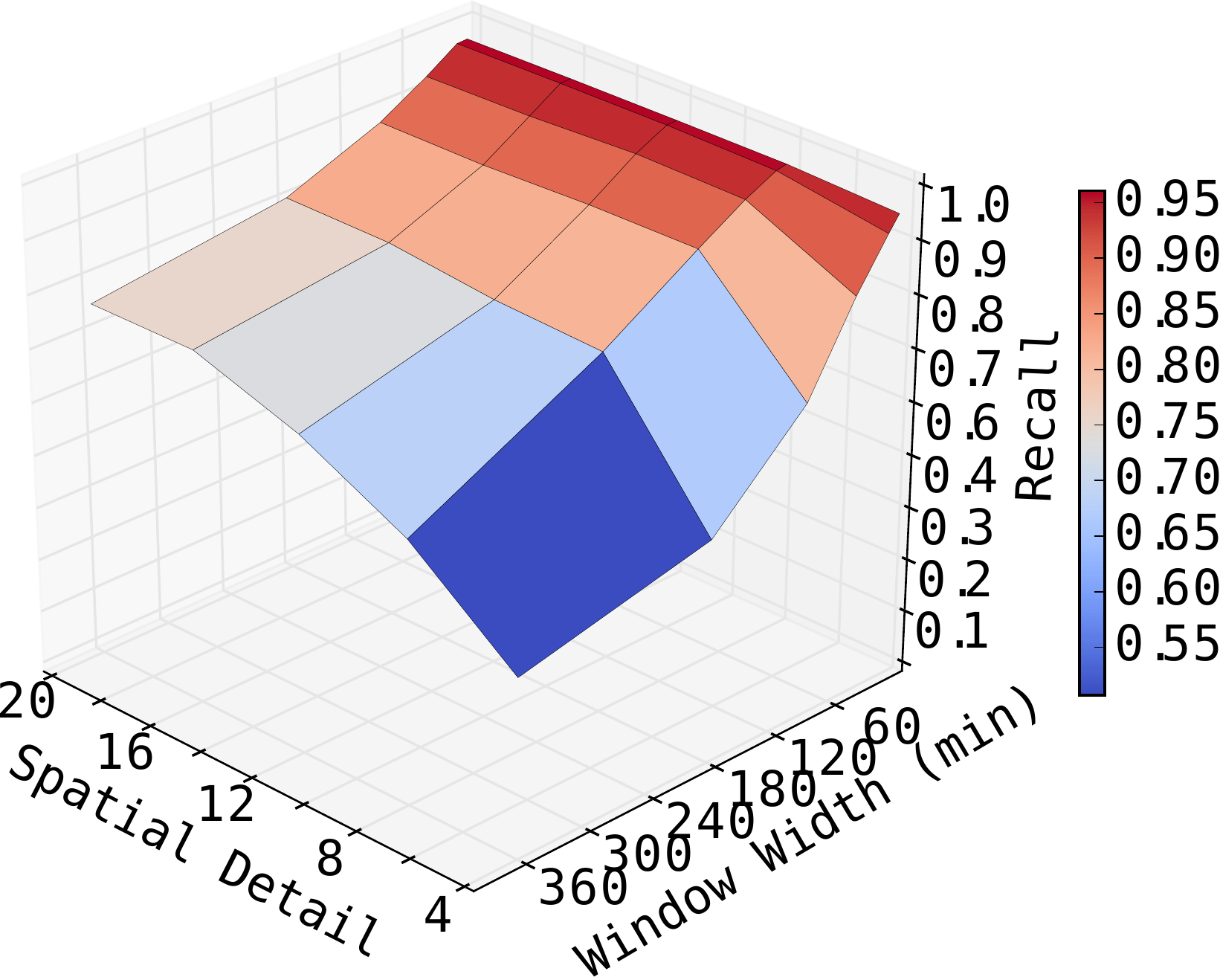}
 		\caption{Recall}
		\label{fig:1-1smrecall}
	\end{subfigure}
  	\hspace{0.0001\linewidth}
	\begin{subfigure}{0.24\linewidth}
  		\includegraphics[width=\linewidth]{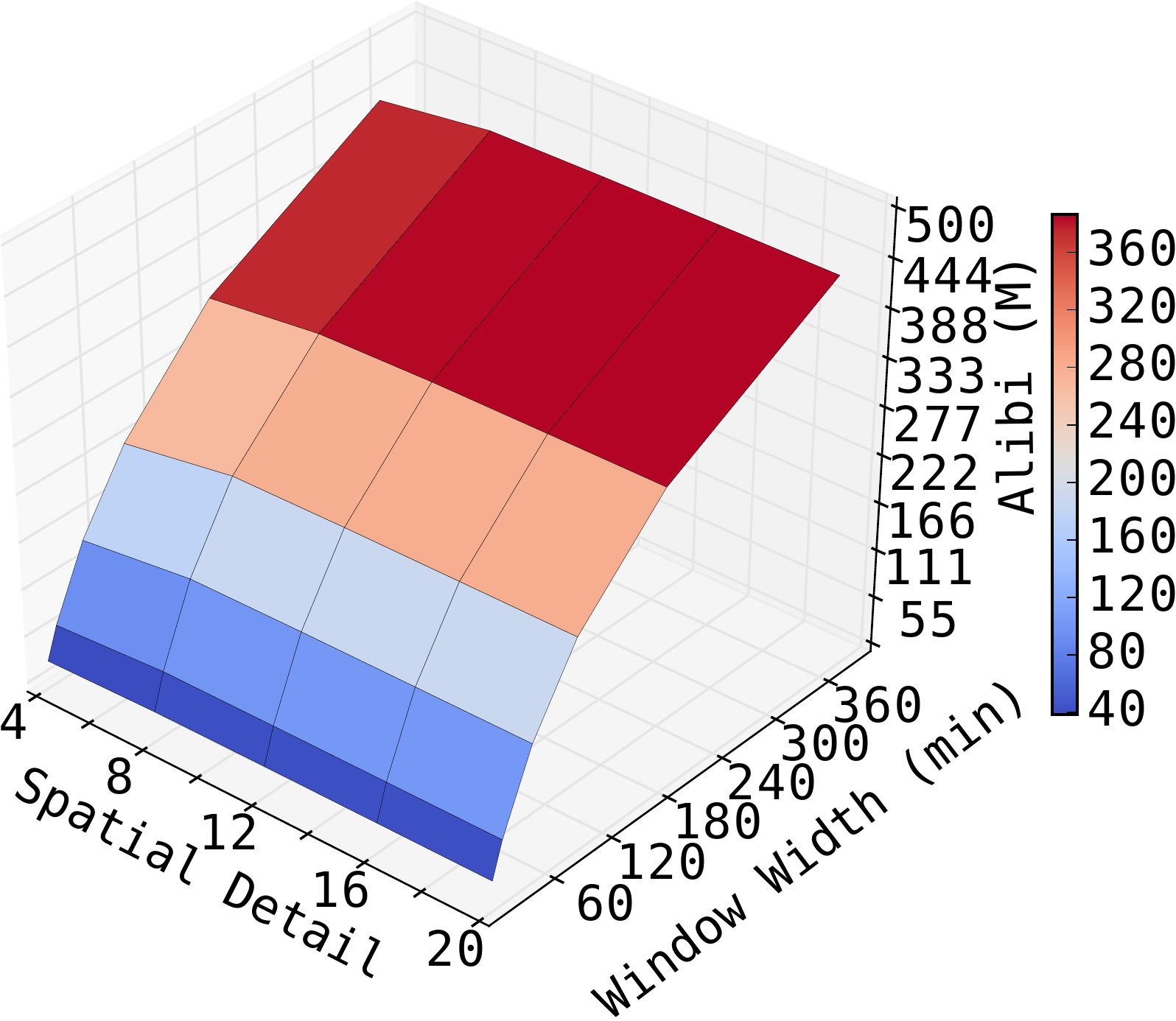}
  		\caption{\# of alibi pairs}
		\label{fig:1-1smalibi}
	\end{subfigure}
	\hspace{0.0001\linewidth}
	\begin{subfigure}{0.24\linewidth}
  		\includegraphics[width=\linewidth]{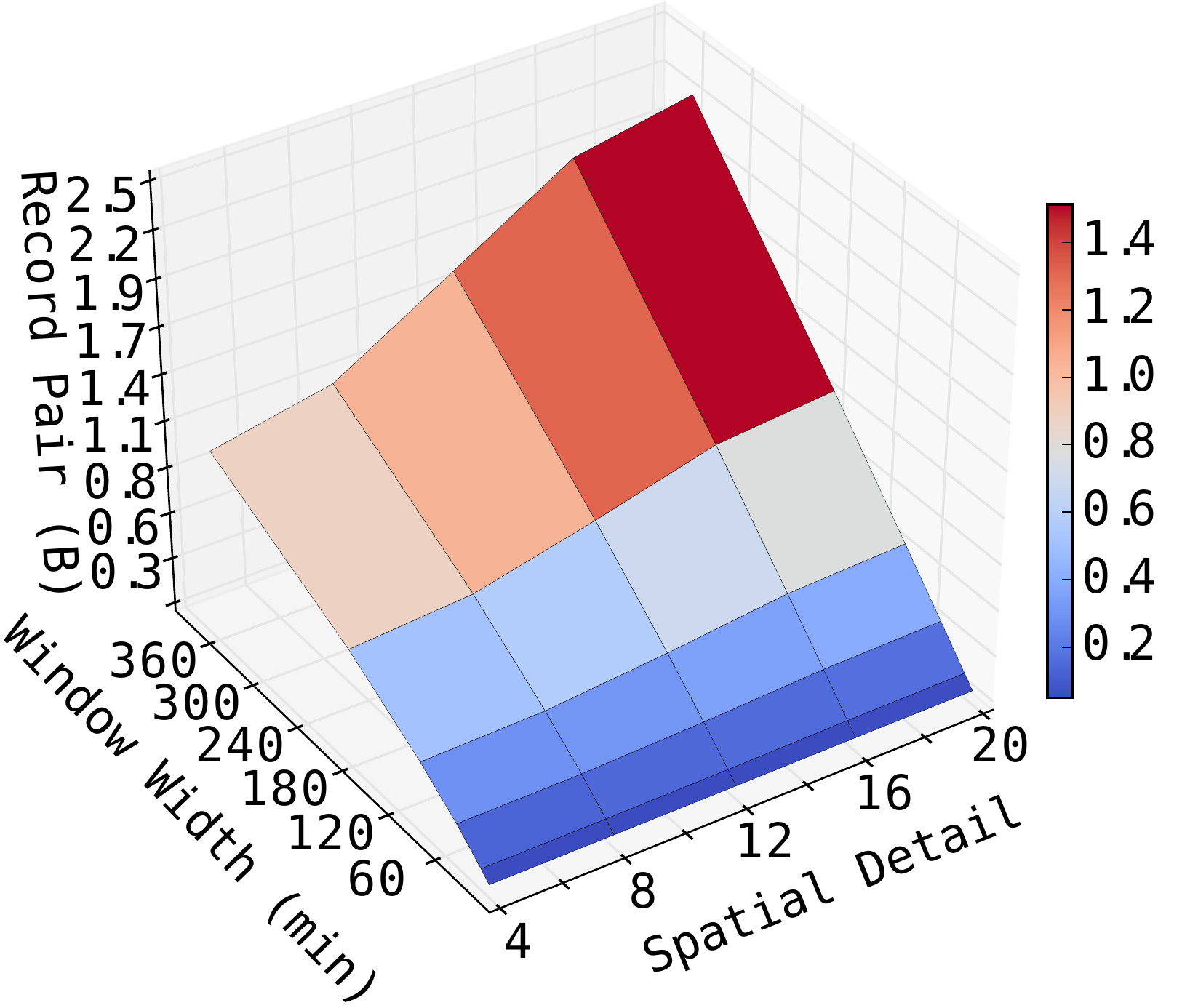}
  		\caption{\# of event comparisons}
		\label{fig:1-1smevent}
	\end{subfigure}
	\caption{Effect of the spatio-temporal level -- SM}
	\label{fig:1-1sm}
\end{figure*}
\begin{figure}
\centering
\includegraphics[width=0.75\linewidth]{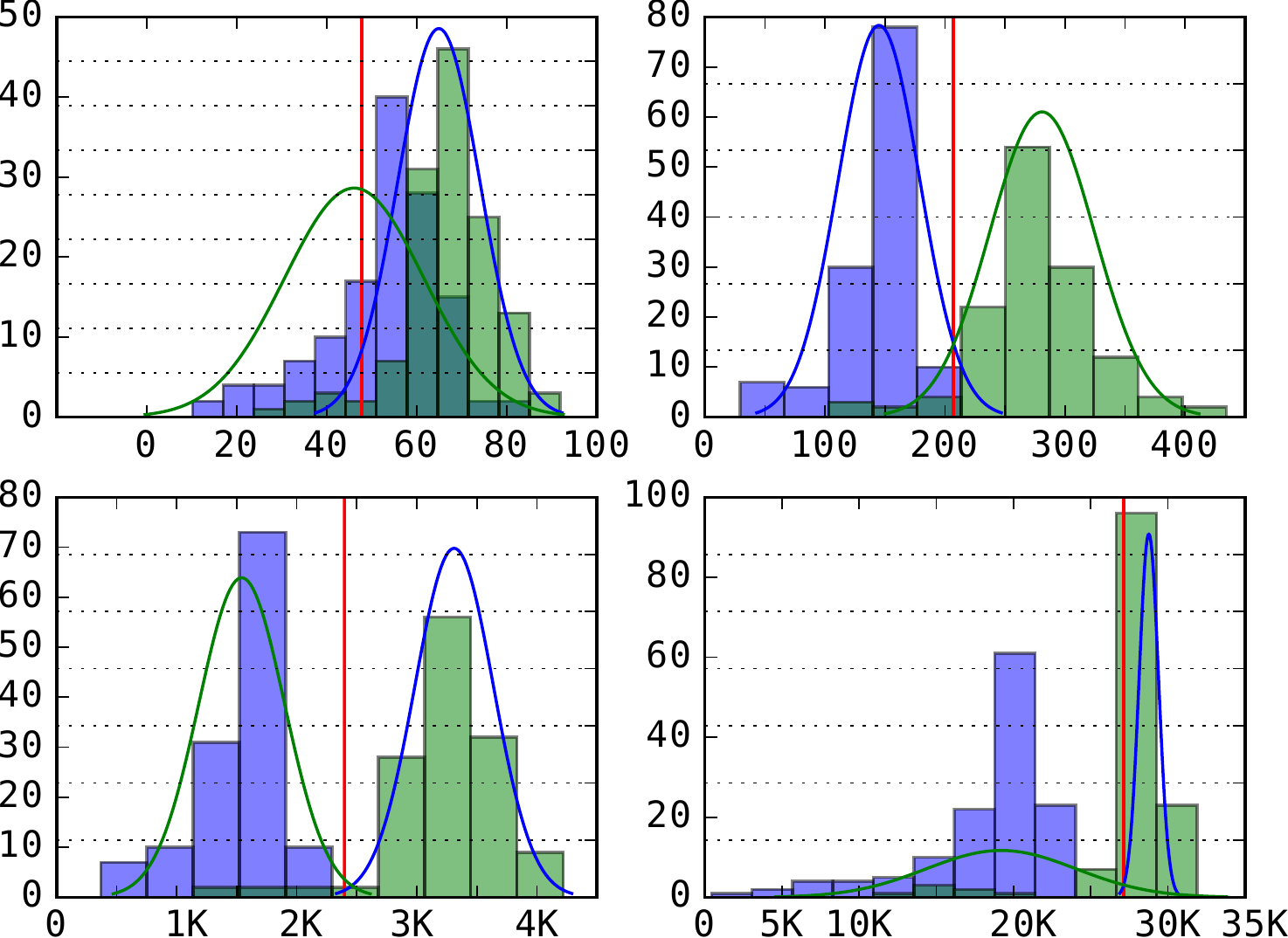}
\caption{Similarity score histograms}% for different spatial details}
\label{fig:split}
\end{figure}
\begin{figure*}[ht]
	\begin{subfigure}{0.24\linewidth}
		\includegraphics[width=\linewidth]{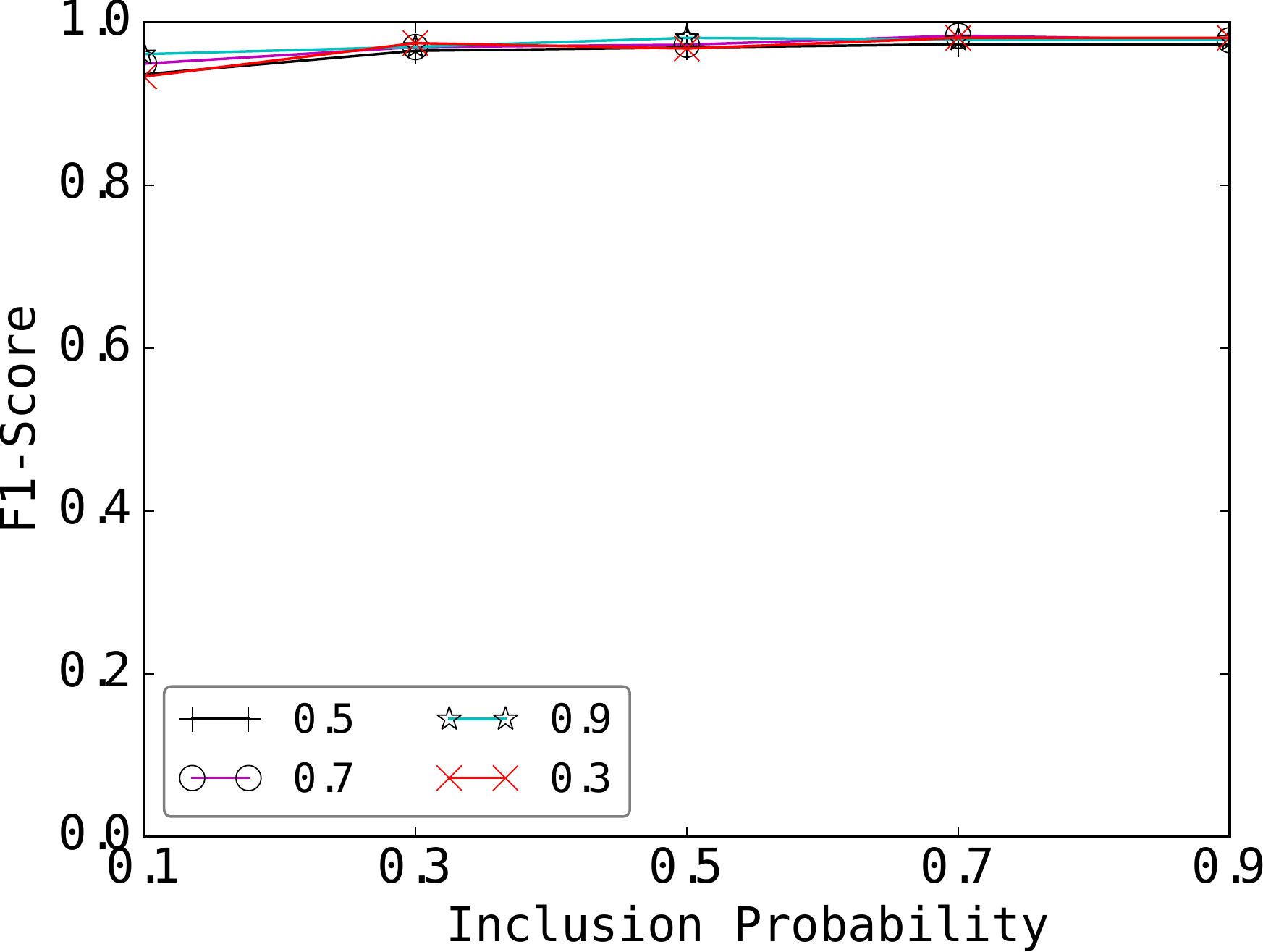}
 		\caption{F1-Score -- Cab}
		\label{fig:1-2cabprecision}
	\end{subfigure}
	\hspace{0.0001\linewidth}
  	\begin{subfigure}{0.24\linewidth}
		\includegraphics[width=\linewidth]{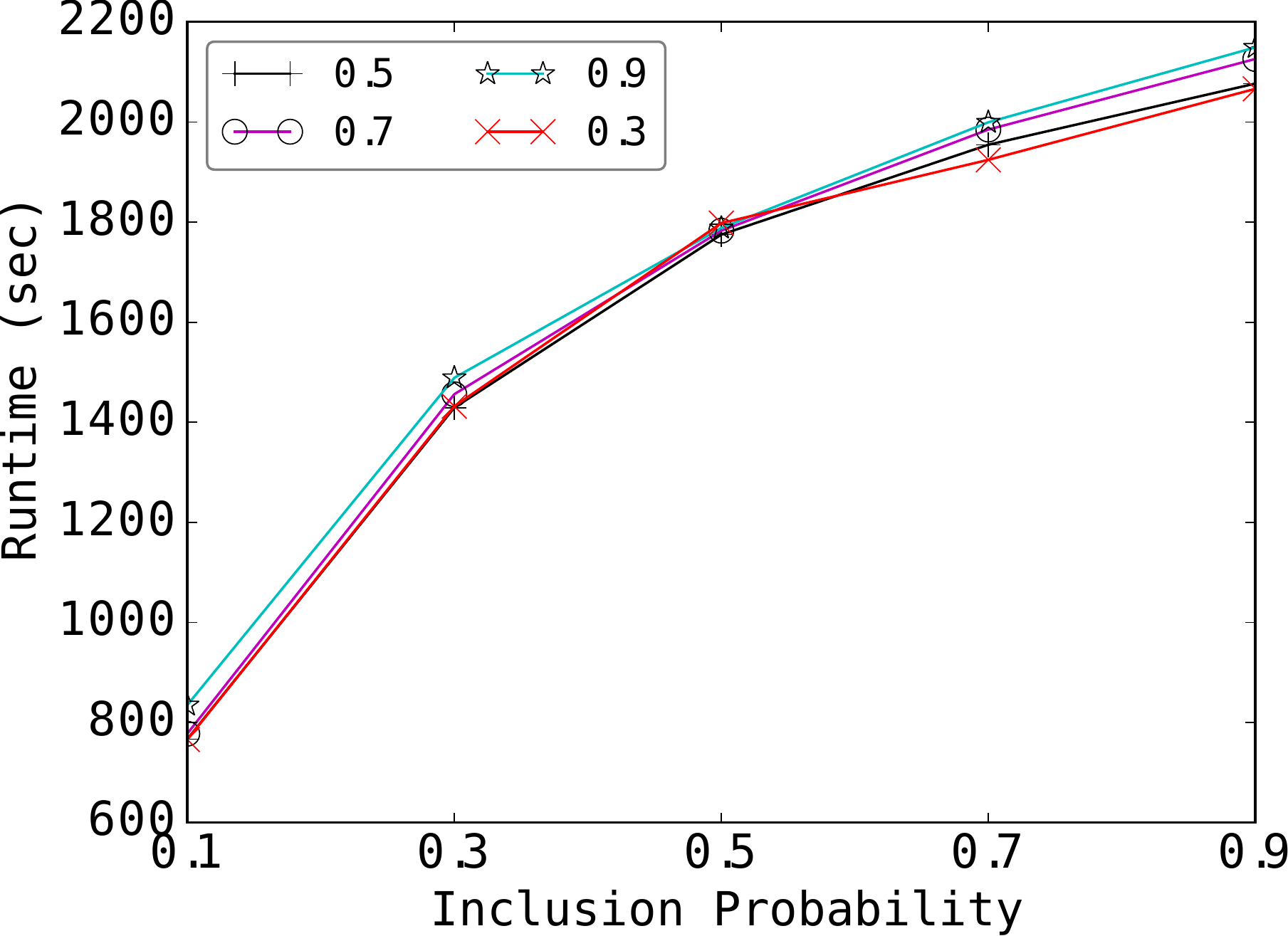}
 		\caption{Runtime -- Cab}
		\label{fig:1-2cabrecall}
	\end{subfigure}
  	\hspace{0.0001\linewidth}
	\begin{subfigure}{0.24\linewidth}
  		\includegraphics[width=\linewidth]{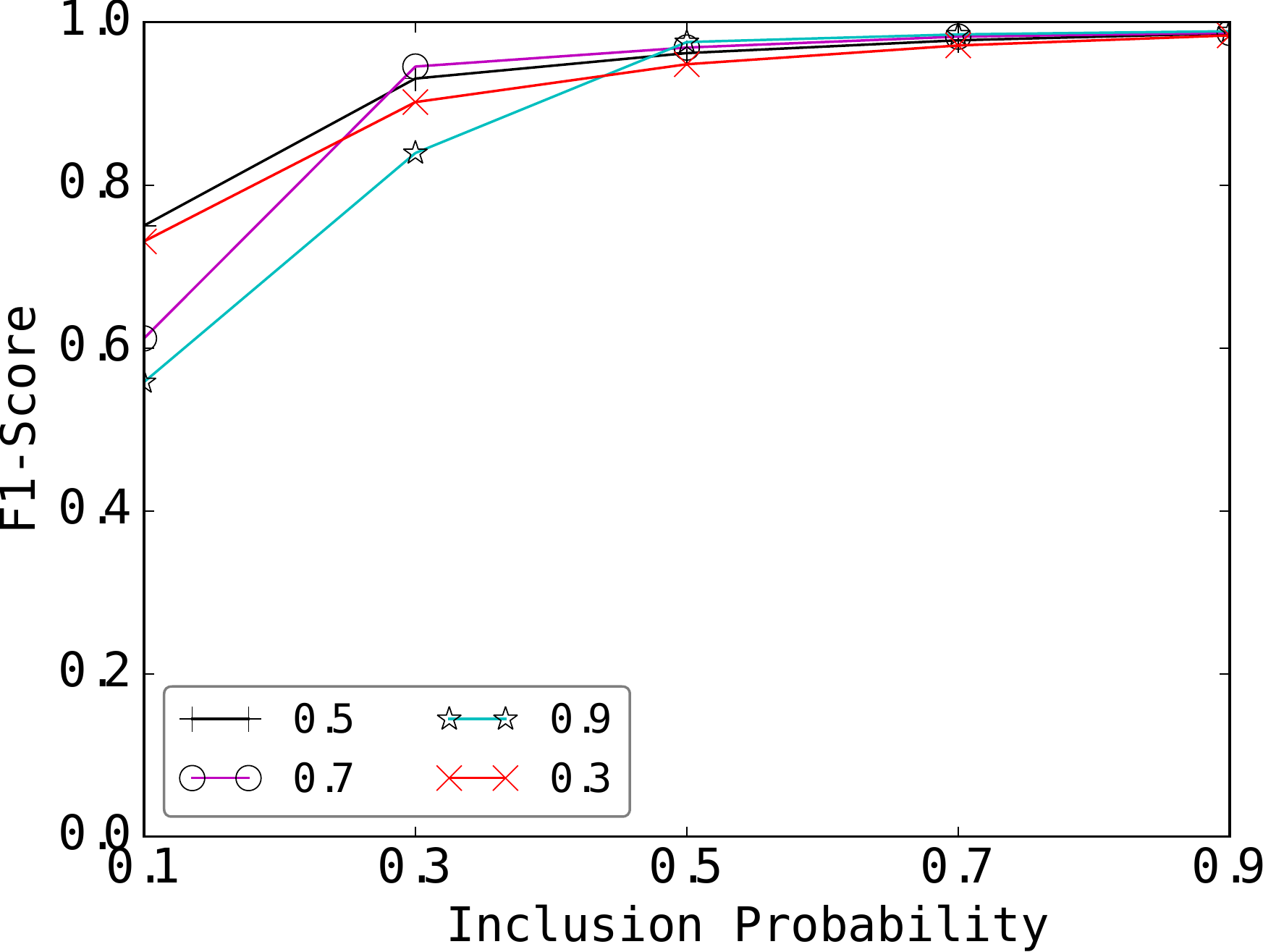}
  		\caption{F1-Score -- SM}
		\label{fig:1-2smprecision}
	\end{subfigure}
	\hspace{0.0001\linewidth}
	\begin{subfigure}{0.24\linewidth}
  		\includegraphics[width=\linewidth]{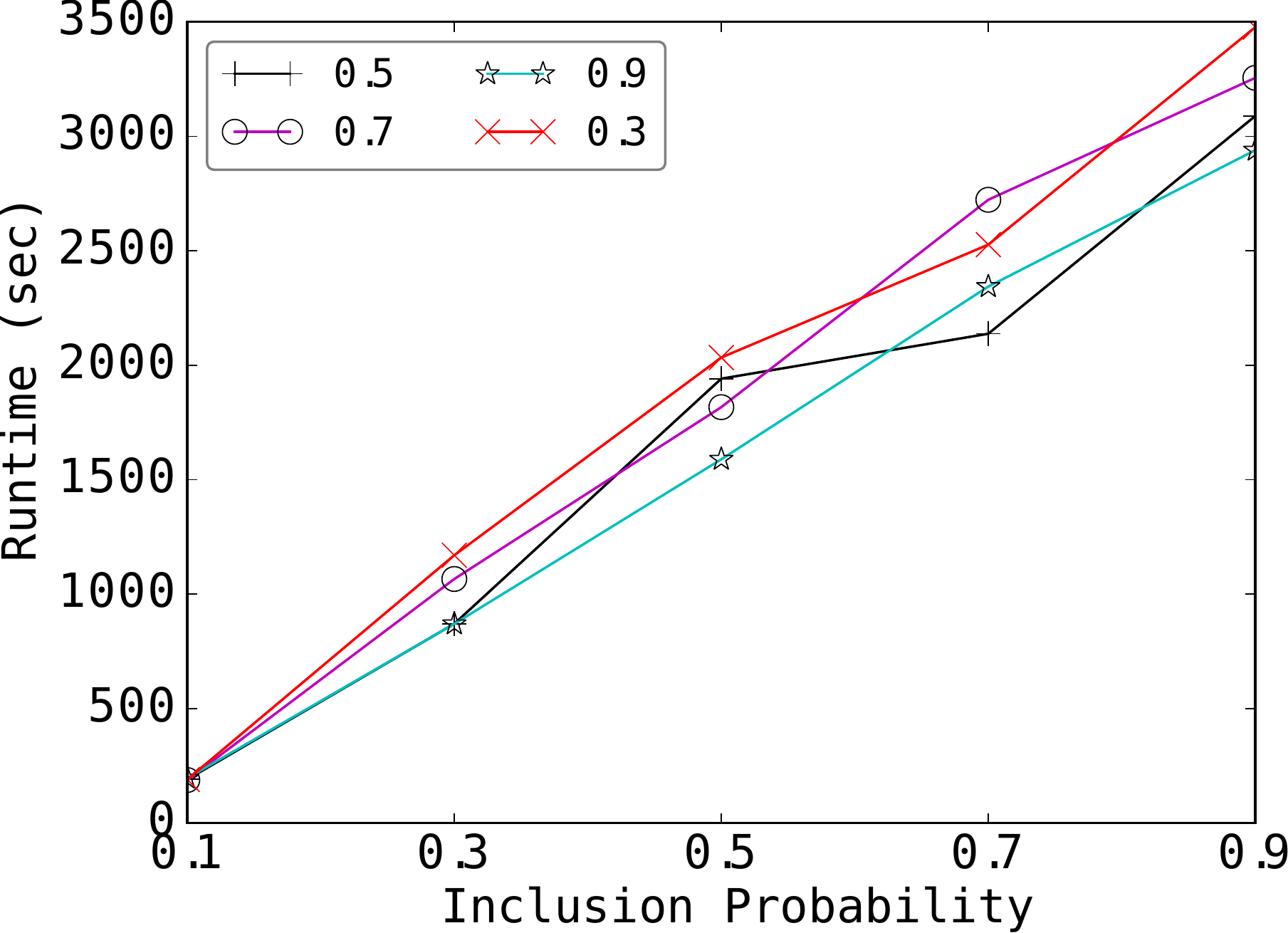}
  		\caption{Runtime -- SM}
		\label{fig:1-2smrecall}
	\end{subfigure}
	\caption{F1-Score and Runtime as a function of the inclusion probability (for different entity intersection ratios)}
	\label{fig:exp1-2}
\end{figure*}
\begin{figure*}[ht]
	\begin{subfigure}{0.24\linewidth}
		\includegraphics[width=\linewidth]{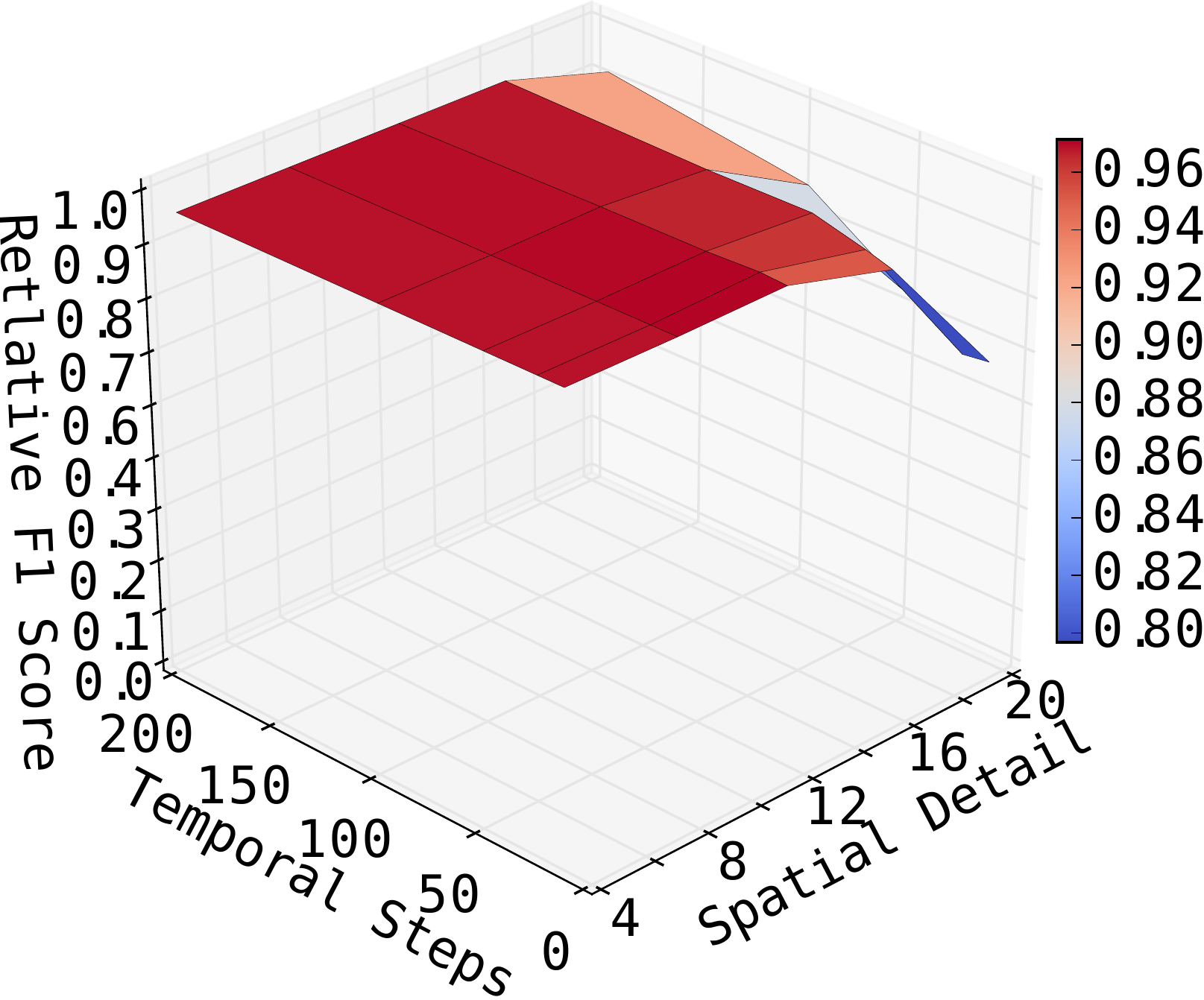}
 		\caption{$F1$-Score -- Cab}
		\label{fig:2-1cabf1}
	\end{subfigure}
	\hspace{0.0001\linewidth}
  	\begin{subfigure}{0.24\linewidth}
		\includegraphics[width=\linewidth]{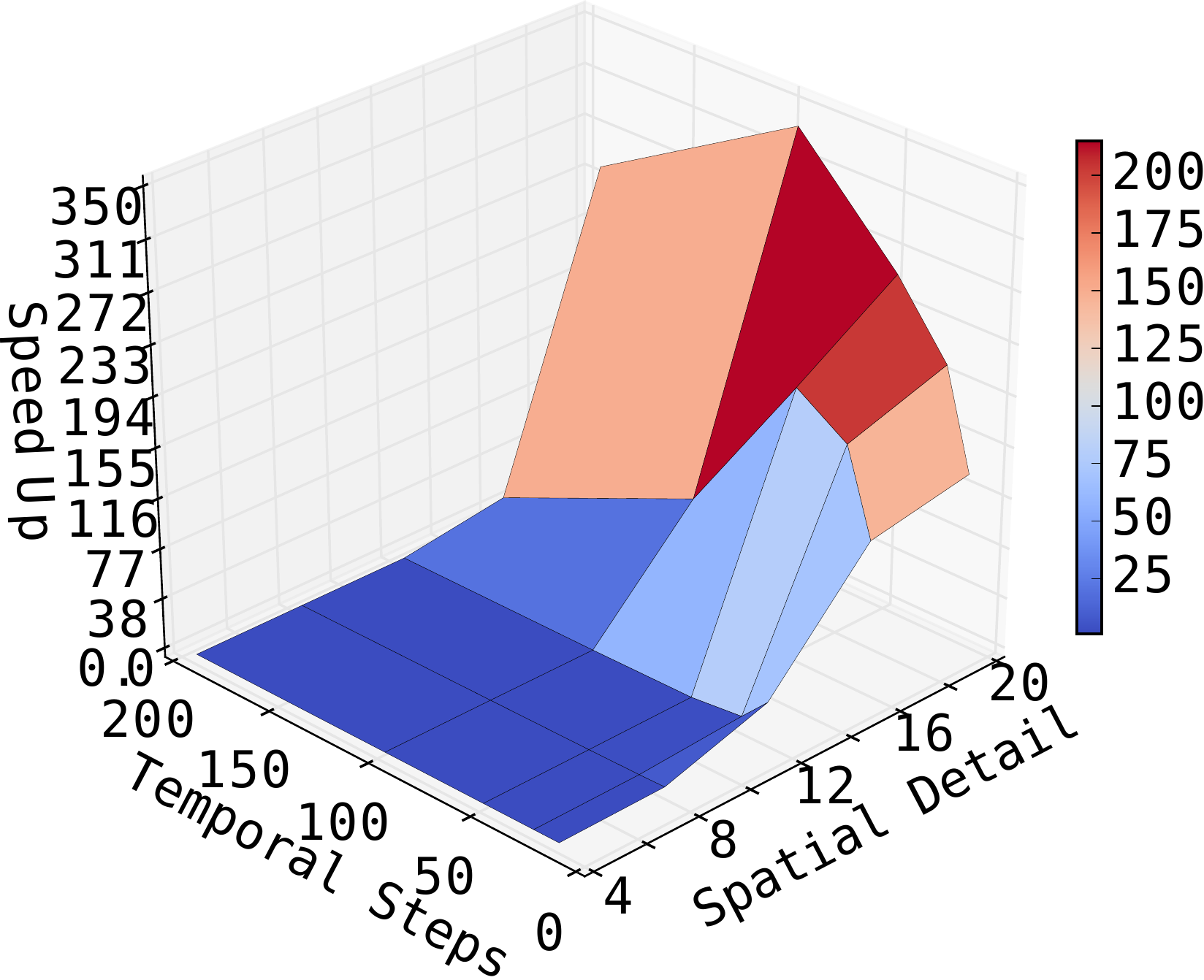}
 		\caption{Speed-up -- Cab}
		\label{fig:2-1cabspeed}
	\end{subfigure}
  	\hspace{0.0001\linewidth}
	\begin{subfigure}{0.24\linewidth}
  		\includegraphics[width=\linewidth]{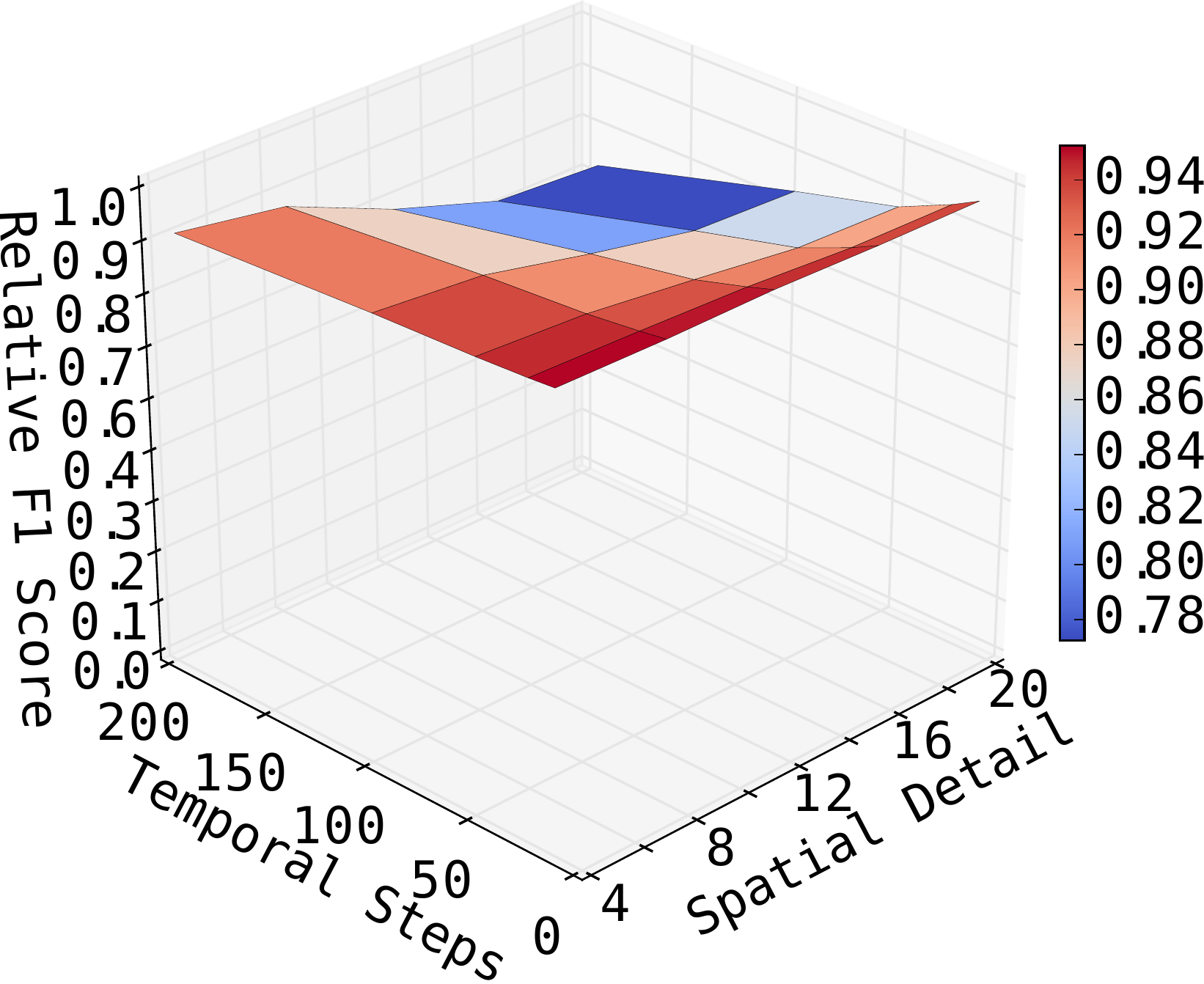}
  		\caption{$F1$-Score -- SM}
		\label{fig:2-1smf1}
	\end{subfigure}
	\hspace{0.0001\linewidth}
	\begin{subfigure}{0.24\linewidth}
  		\includegraphics[width=\linewidth]{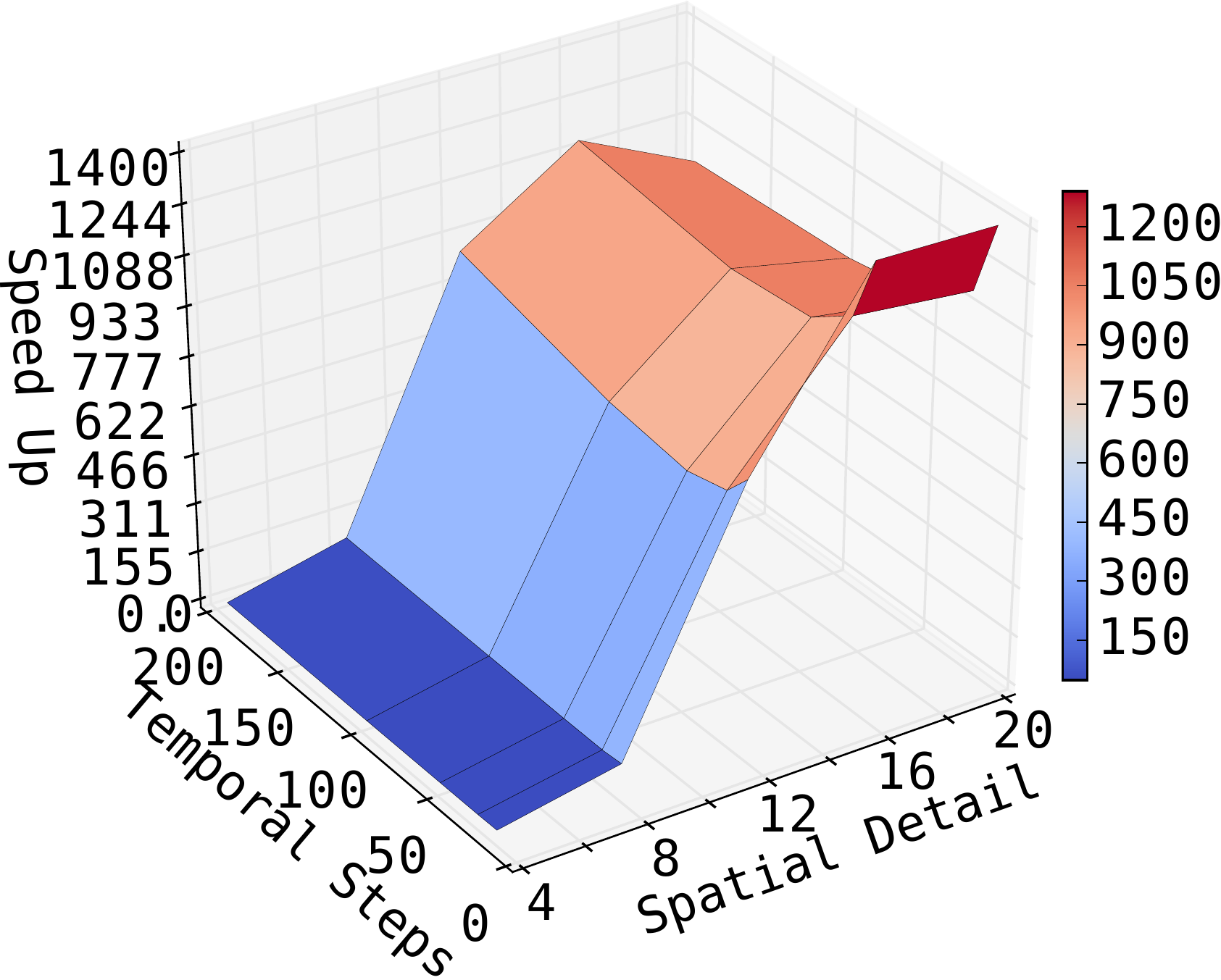}
  		\caption{Speed-up -- SM}
		\label{fig:2-1smspeed}
	\end{subfigure}
	\caption{LSH accuracy and speed-up as a function of the spatial level and temporal step size}
	\label{fig:exp2-1}
\end{figure*}

\begin{figure}[ht]

%	\begin{subfigure}{0.24\linewidth}
%		\includegraphics[width=\linewidth]{figures/2-2-cab-f1.pdf}
% 		\caption{$F1$-Score -- Cab}
%		\label{fig:2-2cabf1}
%	\end{subfigure}
%	\hspace{0.0001\linewidth}
  	\begin{subfigure}{0.48\linewidth}
		\includegraphics[width=\linewidth]{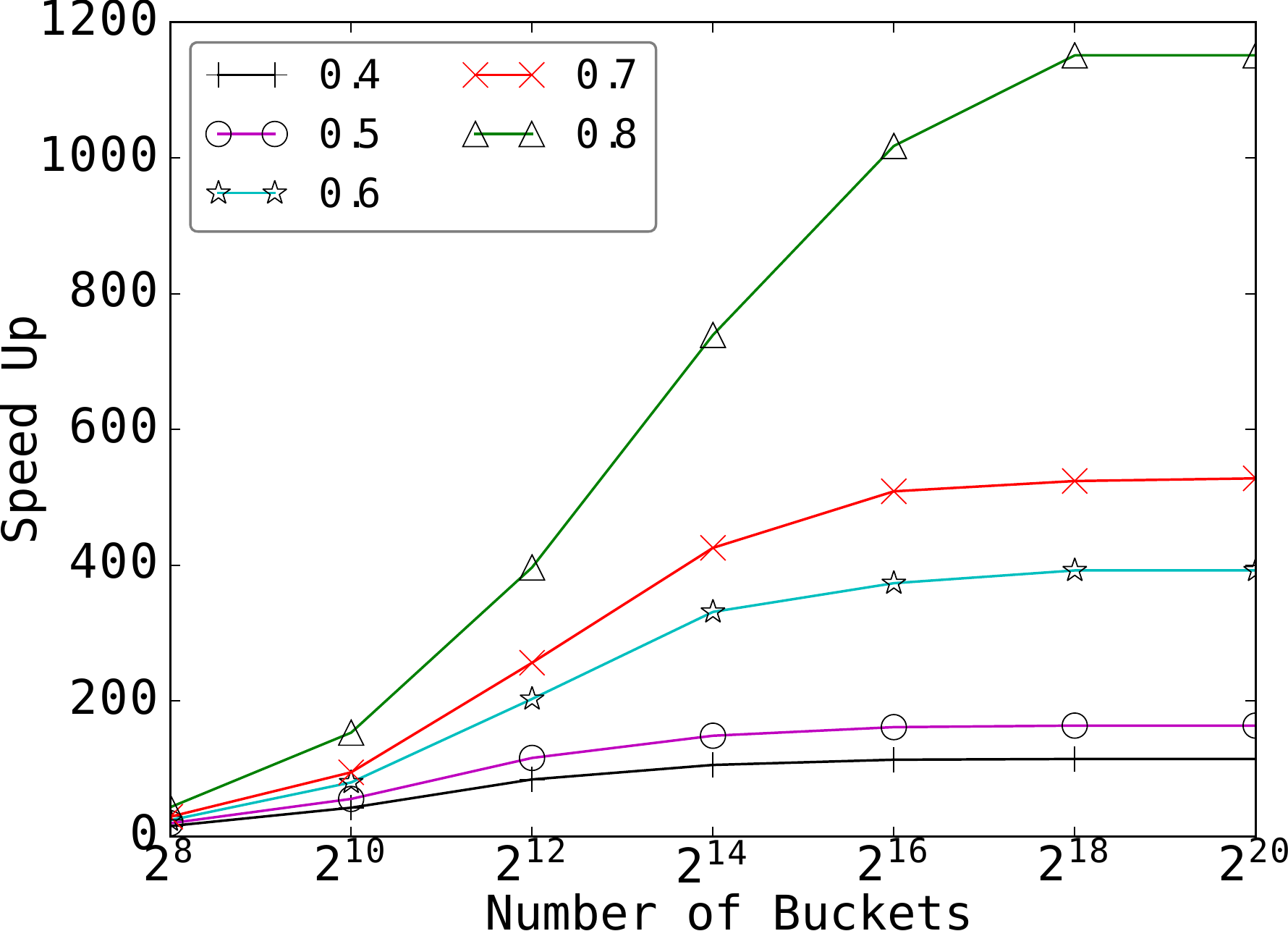}
 		\caption{Speed-up -- Cab}
		\label{fig:2-2cabspeed}
	\end{subfigure}
  	%\hspace{0.0001\linewidth}
	%\begin{subfigure}{0.24\linewidth}
  %		\includegraphics[width=\linewidth]{figures/2-2-sm-f1.pdf}
  %		\caption{$F1$-Score -- SM}
%		\label{fig:2-2smf1}
%	\end{subfigure}
	\hspace{0.0001\linewidth}
	\begin{subfigure}{0.48\linewidth}
  		\includegraphics[width=\linewidth]{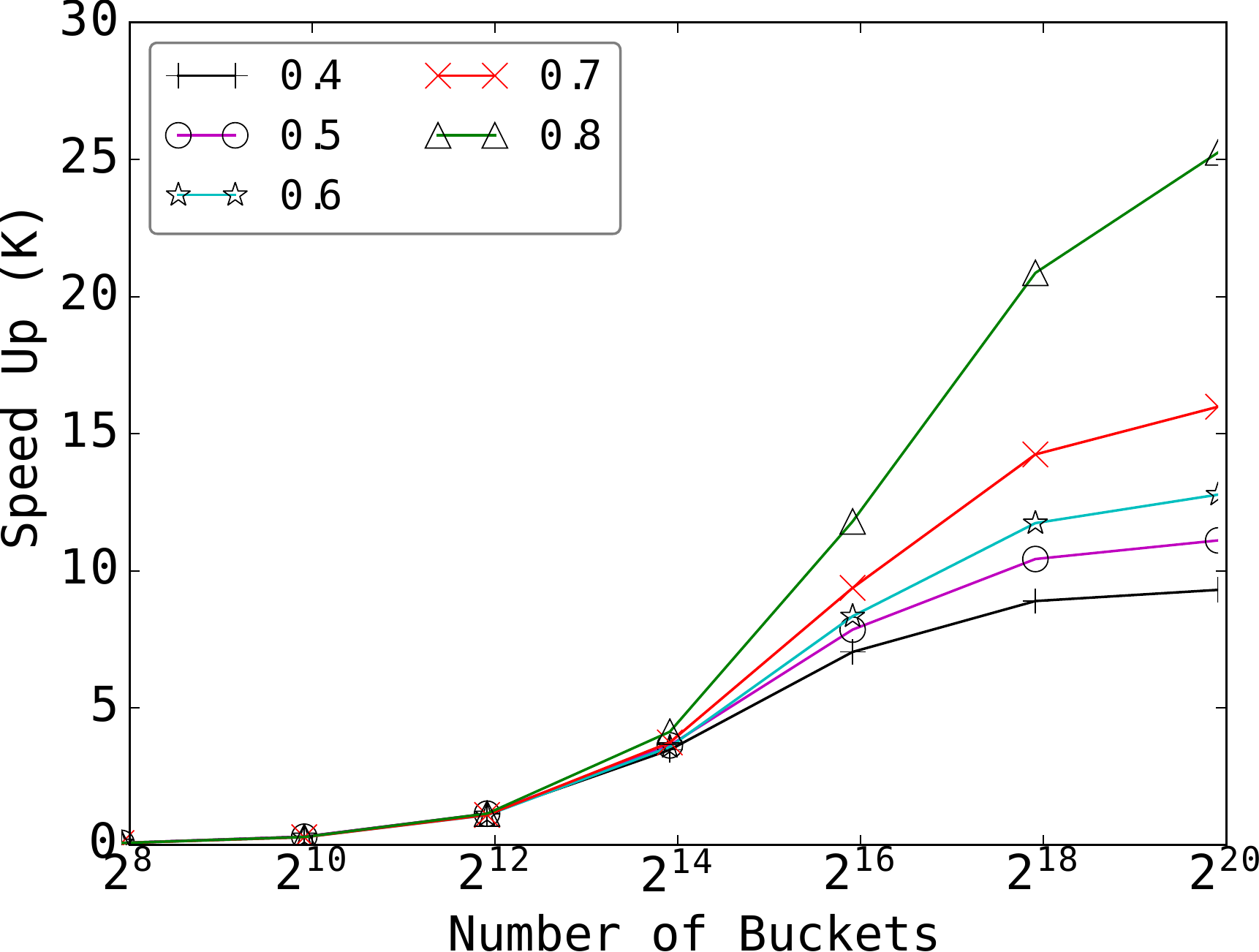}
  		\caption{Speed-up -- SM}
		\label{fig:2-2smspeed}
	\end{subfigure}
	\caption{Speed-up as a function of the bucket size}
	\label{fig:exp2-2}
\end{figure}
\noindent The first set of linkage experiments is performed using the \emph{Cab} dataset which contains mobility traces of approximately $530$ taxis collected over $24$ days (from 2008-05-17 to 2008-06-10) in San Francisco. It has $11,073,781$ records. 
The second set of experiments is performed on linking \emph{Social Media (SM)} data which contain publicly shared records with location information from Foursquare and Twitter. This dataset contains around $5$ million records: $2,266,608$ records from $197,474$ Twitter users and $2,987,747$ records from $276,796$ Foursquare users, distributed over the globe. The dataset spans $26$ days from 2017-10-03 to 2017-10-29. 
In the experimental evaluation we use only time, lat-long and anonymized user-id, and remove all other features.

In each setup (either \emph{Cab} or \emph{SM}), we sample two, possibly overlapping, subsets from each dataset, and link these subsets to each other. To control the experimental setup, we use two parameters during this sampling: %while generating the subsets: 
\emph{entity intersection ratio} and \emph{record inclusion probability}. 
Intuitively, it is unlikely to have the entities from one dataset to be a subset of the other. Therefore, using the entity intersection ratio as a parameter, we control the ratio of the number entities that are common in both datasets to the number of all entities in the smaller dataset. As we show later, this parameter is critical to observe the behavior of SLIM in the presence of false positives. Once the entities are finalized, we also downsample the records from the datasets. This is to address the common case in real life that two location-based services are not always used synchronously in practice and different services might have different usage frequencies. A record of an entity is included in a dataset with the \emph{record inclusion probability}. Higher probability implies denser datasets.

The default values for the entity intersection ratio, the record inclusion probability and the parameter $b$ from Equation~\ref{eq:sim} are equal to $0.5$. We picked $0.5$ as the default, because it is the median of all values in our experimental setup. The spatial detail at the mobility history leaves are controlled using the cell levels of S2. A higher level indicates more spatial detail. The default value for the spatial level is $12$, and the default temporal window width is $15$ minutes. When a mobility dataset is spatially dense, using lower granularity in spatial detail results in too few spatial grid cells. Likewise, for wider temporal windows, one could expect all entities to visit all spatial grid cells. On the other hand, as we show in our experiments, using a higher spatial level of detail does not effect the accuracy but can harm the performance after a certain point. To avoid the adverse effect of entities with too small number of records after downsampling, we ignore an entity if it does not have more than $5$ records. To identify the alibi threshold, we set the maximum movement speed of an entity to 2 km/minute and multiply this constant with the temporal window width. For this value, we took speed-limit of US highways into consideration.

For sections 5.2, 5.3 and 5.4 we link the datasets sampled using default parameters. For the Cab datasets we have two datasets, each with $265$ entities, $133$ of them are in common. Average number of events per entity in these dataset is $10,700$. Similarly, for the SM dataset we have two datasets each with around $30,000$ entities, and $15,000$ of them are in common. Average number of events per entity in these datasets is $12$.

\subsection{Accuracy}
\noindent In this section, we first study precision and recall as a composite function of the spatio-temporal level. We also look at the number of alibi entity pairs and the number of pairwise record comparisons to better understand SLIM's behavior. Next, we study $F1$-Score and running time as a function of the record inclusion probability.

\subsubsection{Effect of the Spatio-Temporal Level}
\noindent Figures~\ref{fig:1-1cab} and~\ref{fig:1-1sm} plot precision, recall, alibi pairs and number of record comparisons as a function of the spatio-temporal level for the Cab and SM datasets, respectively. In all figures, the $x$-axis shows the spatial detail, the $y$-axis shows the width of the window in minutes, and the $z$-axis shows the measure.

Figures~\ref{fig:1-1cabprecision}~and~\ref{fig:1-1cabrecall} show precision and recall for the Cab dataset. We observe that both measures increase with spatial detail. This is because when the spatial detail increases, the distance calculation becomes more accurate. 
After spatial level $12$, $F1$-Score %precision and recall values 
becomes greater than $0.95$. However, for window width, after $90$ minutes, while recall remains high, precision decreases dramatically. For spatial detail $20$, when the window size is $15$ minutes, perfect precision is reached, while for window size $360$ minutes the precision is $0.56$. The decrease in precision is steeper for spatial detail $20$ than spatial detail $16$, but for the same data point recall is higher for spatial detail $20$ than $16$. The reason behind this is that, since the records in the same time-location bins are aggregated, using large time-windows makes it harder to distinguish entities from each other. When the level of detail is low, both spatially and temporally, the variance of entity pair scores is decreasing. To observe this behavior better, Figure~\ref{fig:split} shows detected stop threshold values (red lines), fit GMM models (blue and green curves), and distribution of true positive and false positive links for spatial detail values $4$, $8$, $12$, and $16$ as a function of similarity scores for window width $90$ minutes. We observe that with increasing spatial detail, grouping true positive links (green bars) and false positive links (blue bars) in two clusters becomes more accurate and a tighter stop threshold value could be identified. By looking at the distances between two components of GMM one could say that stop threshold identification has subpar accuracy for spatial detail values lower than $12$. One could observe this subpar accuracy by looking at the differences between precision behaviors of  Figures~\ref{fig:1-1cabprecision} and~\ref{fig:1-1smprecision}. We observe similar results using Otsu technique~\cite{ref:otsu} and 2-means clustering but those experiments are omitted due to space constraints. For low spatial detail (i.e., $\leq 10$) and high temporal window width (i.e., $\geq 60$ minutes), precision is favored over recall for the Cab dataset but vice-versa for SM.

While spatial level values higher than $12$ have similar 
precision and recall, we observe that increasing spatial detail also increases the number of record comparisons. This is expected, as we discussed in Section~\ref{sec:tune} how to use the trade-off between accuracy and performance to detect the best spatial detail for a given temporal window. When the window size is $15$ minutes, the spatial detail detected by the parameter tuning algorithm is $12$. Figure~\ref{fig:1-1cabevent} shows that for the same window width, increasing spatial detail from $12$ to $20$ increases the number of pairwise record comparisons by $1.14$ times, yet the accuracy stays the same. The gap widens for longer temporal windows. The same figure  
shows $3.15\times$ increase in the number of record comparisons when the window size is increased from $15$ to $360$ minutes, for spatial detail $12$. Yet, the increase is $22\times$ for spatial detail $20$. 

Figure~\ref{fig:1-1sm} shows the same experiment for the SM dataset. Most of the previous observations hold for this dataset as well. There are two additional observations. First, in the Cab dataset the best recall value is reached when $5$-minute windows are used. This is a result of the spatio-temporal density of the Cab dataset, as alibi detection becomes more efficient for narrow windows, resulting in better recall. On the other hand, in the SM dataset, the best recall is reached for $15$-minute windows. This is expected, because at one extreme very small temporal windows require services to be used synchronously to collect evidence for linkage. Another observation is that, as SM dataset has lower spatio-temporal skew, to detect alibis one needs to use larger temporal windows. 

\subsubsection{Sensitivity to the Workload Parameters}
\label{sec:sensitivity}
\noindent In this experiment, we link one dataset of each source with the datasets sampled with different entity intersection ratios and record inclusion probabilities. Figure~\ref{fig:exp1-2} plots the $F1$-Score and running time in seconds as a function of record inclusion probability for the Cab and SM datasets, respectively. Different series represent different entity intersection ratios. The Cab dataset has $265$ entities and the SM dataset has around $30,000$ entities. The average number of records for an entity is ranging from $2,100$ to $18,900$ for the Cab dataset and from $10$ to $20$ for the SM dataset.
Figure~\ref{fig:1-2cabprecision} shows the results for the Cab dataset. We observe that all $F1$-Score values are close to $1$, even when average number of records are as low as $2,100$ (inclusion probability $0.1$). Moreover, from Figure~\ref{fig:1-2cabrecall}, we observe that the running time is sub-linear with average number of records, which is a result of aggregation on mobility histories. These results validate robustness of SLIM, as $F1$-Score is not effected by the increasing number of records and the system scales linearly.

\begin{figure}[ht]
  	\begin{subfigure}{0.48\linewidth}
		\includegraphics[width=\linewidth]{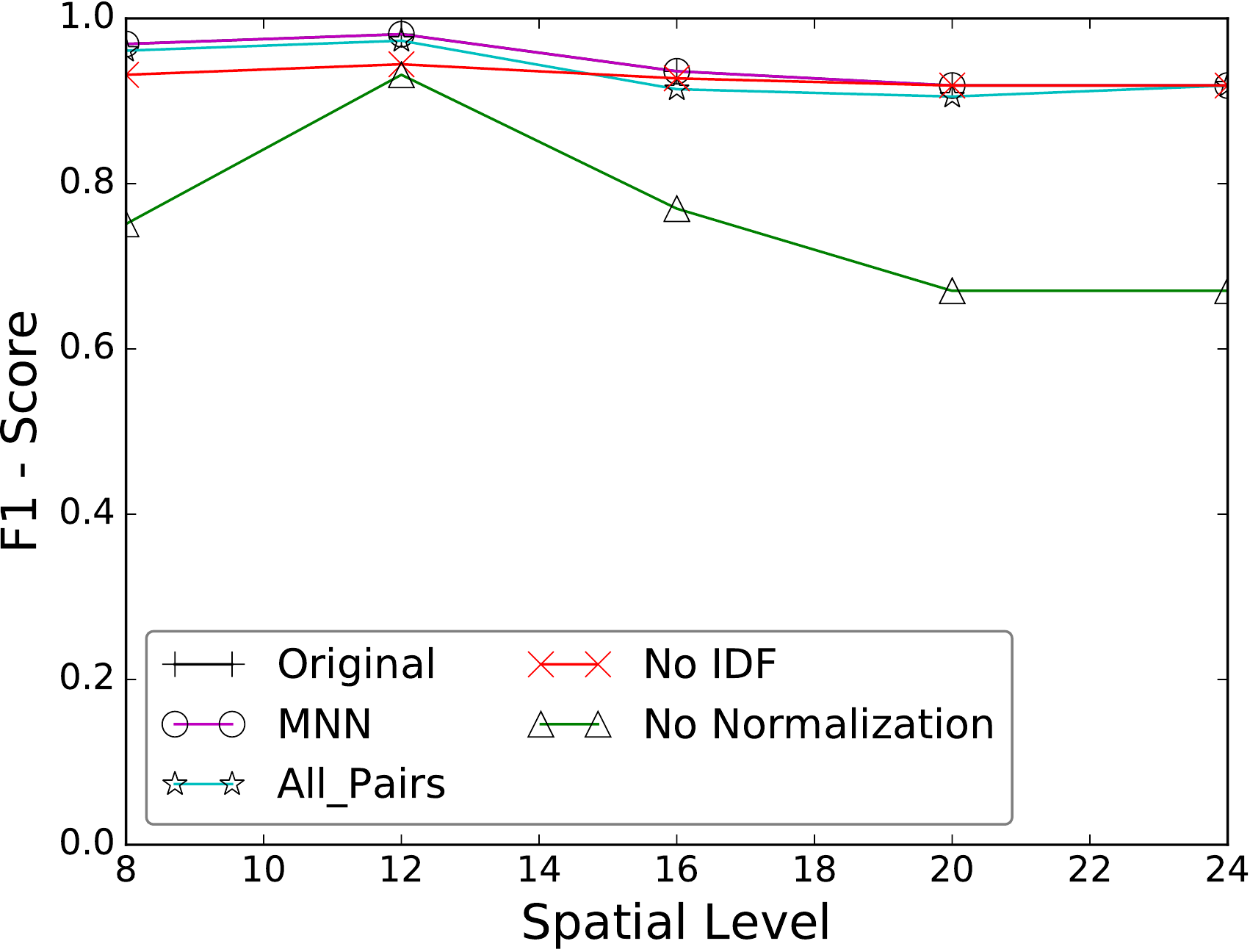}
 		\caption{F1-Score Spatial Level}
		\label{fig:ablation_spatial}
	\end{subfigure}
	\hspace{0.0001\linewidth}
	\begin{subfigure}{0.48\linewidth}
  		\includegraphics[width=\linewidth]{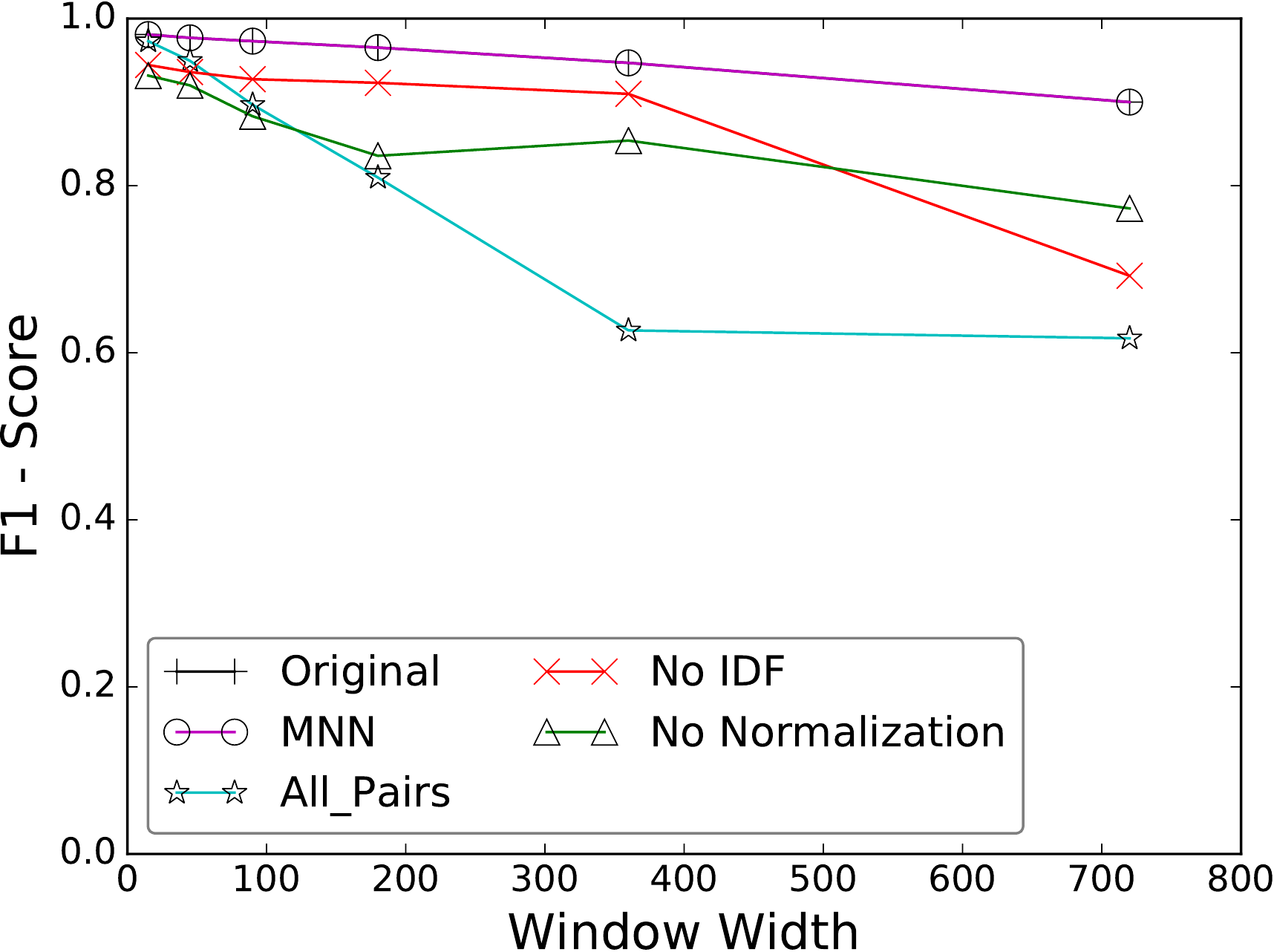}
  		\caption{F1-Score Window Width}
		\label{fig:ablation_temporal}
	\end{subfigure}
	\caption{Ablation Study}
	\label{fig:expAblation}
\end{figure}

Figures~\ref{fig:1-2smprecision}~and~\ref{fig:1-2smrecall} show the results of the same experiment for the SM dataset. Different than the Cab dataset, the effect of the record inclusion probability on the $F1$-Score is more pronounced here. For the entity intersection ratio value of $0.5$, the average number of records is $10$, $F1$-Score is $0.75$. When the average number of records is doubled, we get $0.98$ as the $F1$-Score. This is because the average number of records per entity is already low in the SM dataset and downsampling it to even lower values decreases the number of records that can serve as evidence for linkage. However, we observe that SLIM is able to perform high-accuracy linkage when the average number of records per entity is at least $15$. Independent of the entity intersection ratio, after $15$ records per entity, the $F1$-Score of SLIM is greater than $0.9$. Similar to Cab dataset, the running time of SLIM is linear with the input size for SM dataset.

\begin{figure*}[ht]
	\begin{subfigure}{0.24\linewidth}
		\includegraphics[width=\linewidth]{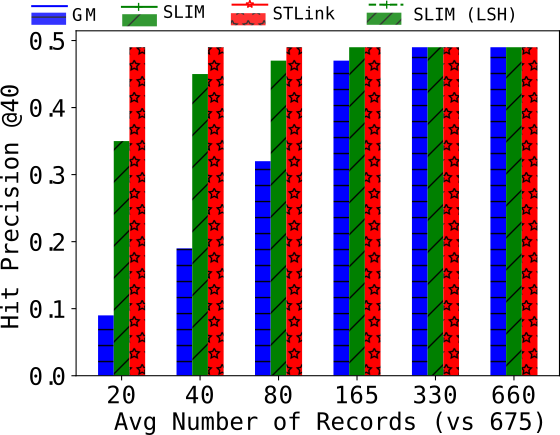}
 		\caption{Hit precision @40}
		\label{fig:3-all-hit}
	\end{subfigure}
	\hspace{0.0001\linewidth}
  	\begin{subfigure}{0.24\linewidth}
		\includegraphics[width=\linewidth]{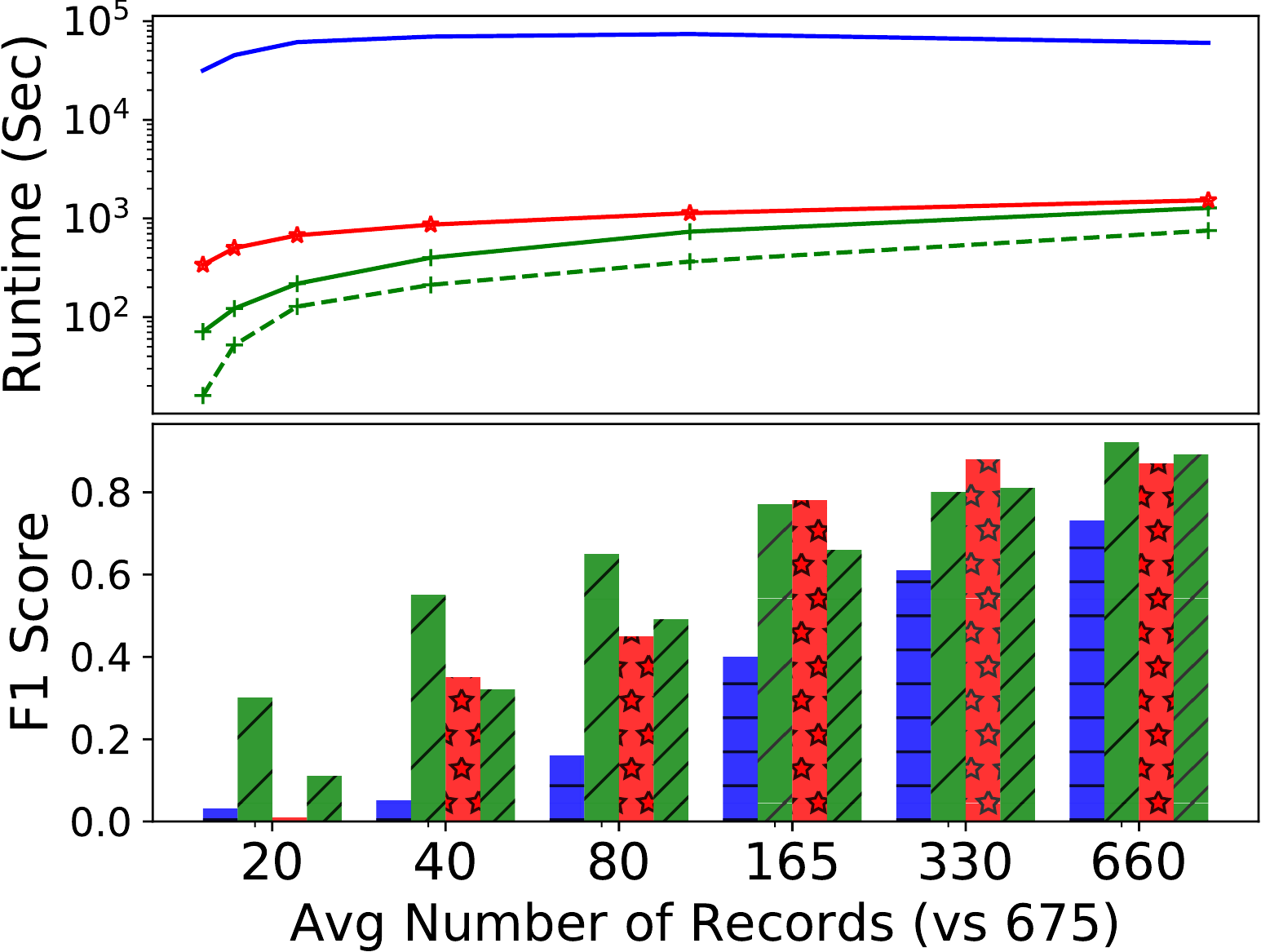}
 		\caption{$F1$-Score}
		\label{fig:3-all-f1}
	\end{subfigure}
  	\hspace{0.0001\linewidth}
	\begin{subfigure}{0.24\linewidth}
  		\includegraphics[width=\linewidth]{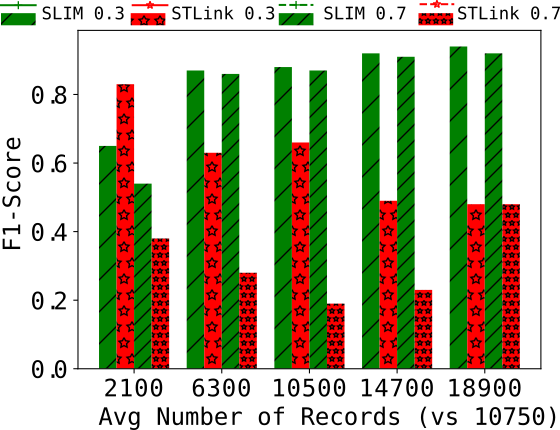}
  		\caption{$F1$-Score}
		\label{fig:3-stlink-f1}
	\end{subfigure}
	\hspace{0.0001\linewidth}
	\begin{subfigure}{0.24\linewidth}
  		\includegraphics[width=\linewidth]{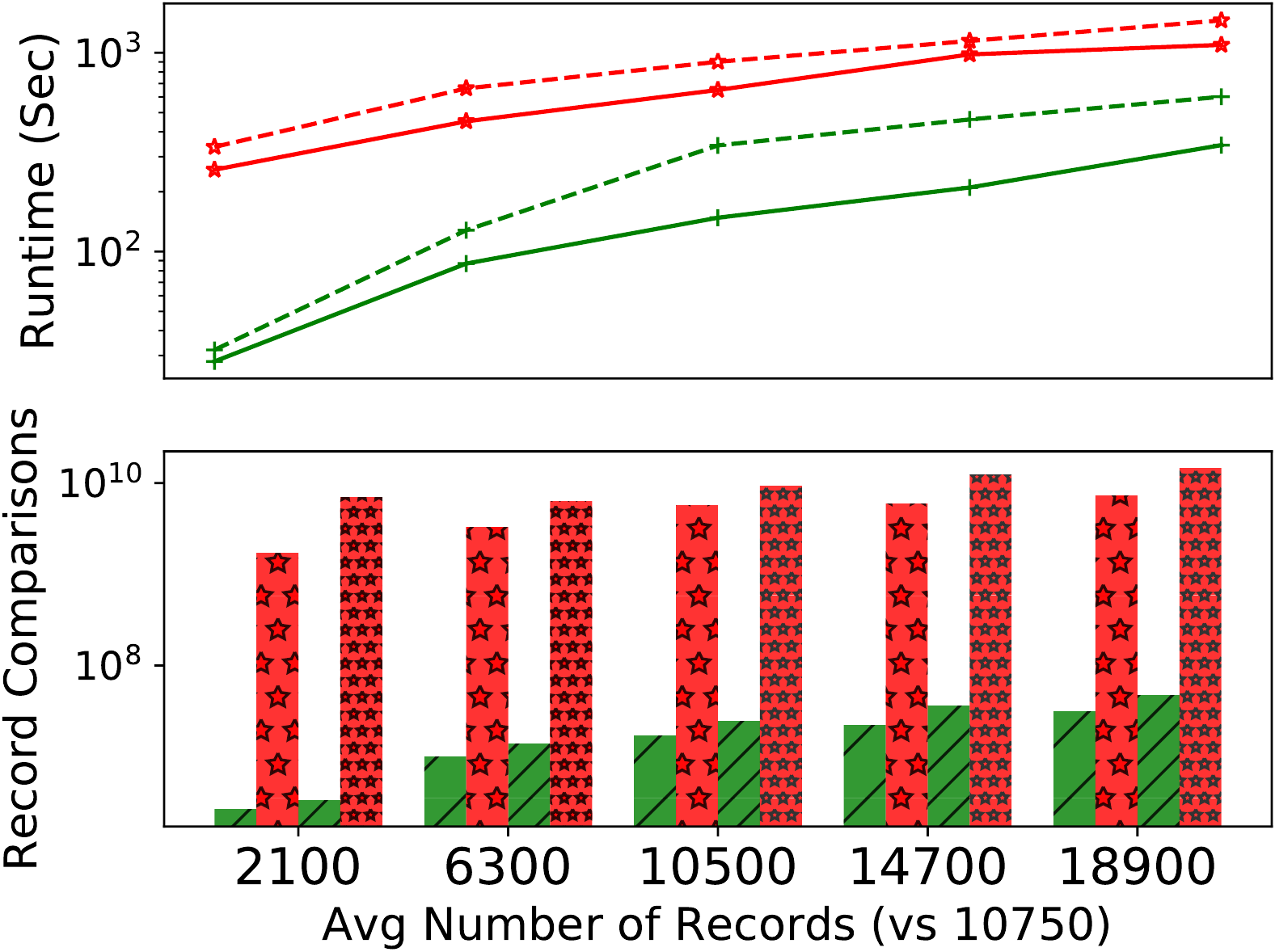}
  		\caption{Record comparisons}
		\label{fig:3-stlink-speedup}
	\end{subfigure}
	\caption{Comparison with existing work (Sub-figures a and b, and c and d are sharing their legends)}
	\label{fig:3-comparisons}
\end{figure*}
\subsection{Scalability}
\noindent In this set of experiments, we study the effect of the LSH on the quality and the scalability of the linkage. The quality is measured using the $F1$-Score relative to that of the brute force linkage. 
Let $F1$-Score where LSH is applied be $F1$-Score$_{lsh}$ and without it $F1$-Score$_{bf}$. Then, relative $F1$-Score equals $F1$-Score$_{lsh}$  / $F1$-Score$_{bf}$.
Likewise, to measure the speed-up, we compute the ratio of the number of pairwise record comparisons without LSH to that of with LSH. 

\subsubsection{Effect of the Spatio-Temporal Level}
\noindent
Figure~\ref{fig:exp2-1} shows the relative $F1$-Score and the speed-up as a composite function of the spatial level and the temporal step size. Recall that we construct a set of dominating cells to act as signatures. This construction is done by querying each mobility history for non-overlapping time windows. The size of each grid cell is defined by the spatial level. The temporal step size represents the number of time windows the query spans. Note that these parameters are different than the spatial level and the window width that is used for the similarity score computation. 
The LSH similarity threshold $t$ is set to $0.6$ and the number of buckets is set to $4096$.

 Figures~\ref{fig:2-1cabf1}~and~\ref{fig:2-1cabspeed} show the relative $F1$-Score and speed-up, respectively, for the Cab dataset. Figure~\ref{fig:2-1cabf1} shows that the $F1$-Score achieved with and without LSH are almost the same when the spatial level is lower than $12$. Similarly, we do not observe any speed-up for these data points. The reason is that the Cab dataset is spatially too dense and consequently dominating grid cells of all entities end up being the same when the spatial detail is low. However, when the spatial detail is increased, we observe that LSH brings $2$ orders of magnitude speed-up by preserving $98\%$ of the $F1$-Score. For the spatial detail value of $16$ and temporal step size of $48$ (this means each dominating grid cell query spans $12$ hours) the speed up reaches $202\times$. The maximum speed-up achieved for this dataset is $332\times$, preserving $86\%$ of the $F1$-Score.

Figures~\ref{fig:2-1smf1}~and~\ref{fig:2-1smspeed} present the same experiment for the SM dataset. While we observe similar behavior for low spatial detail, we also observe that the increase in the speed-up starts earlier and is steeper when the spatial detail is increased. This is because the SM dataset has lower geographic and temporal skew.  If we observe the same data point as we did with the Cab dataset, we observe more pronounced speed-ups: For the spatial detail value of $16$ and temporal step size of $48$, LSH brings $1177\times$ speed-up preserving $91\%$ of the $F1$-Score. Next, we show that the maximum reachable speed-up is much higher when the number of buckets is increased.

\subsubsection{Effect of the LSH Parameters}
\noindent
Figure~\ref{fig:exp2-2} plots
the speed-up as a function of the number of hash buckets. Different series represent different LSH similarity thresholds. We set the spatial detail and temporal step size of the signature calculation to $16$ and $48$, respectively.

Intuitively, $F1$-Score is not effected by the number of buckets. This is because if two entities have at least one identical band in their signatures, they are hashed to the same bucket independent from the number of buckets. Yet, increased number of buckets increases the speed up as the probability of hash collision decreases. Similarly, the LSH similarity threshold affects the relative $F1$-Score. This is because, its smaller values increase the probability of becoming a candidate pair.

We observe the increase in speed-up for the Cab dataset and the SM dataset in Figures~\ref{fig:2-2cabspeed}~and~\ref{fig:2-2smspeed}, respectively. When the number of buckets is set to $2^{18}$ and the similarity threshold is set to $0.6$, the speed-up becomes $380\times$ with a relative $F1$-Score of $0.98$ for the Cab dataset and $11,742\times$ with a relative $F1$-Score of $0.91$ for the SM dataset. Since the number of entities in the SM dataset is much larger compared to the Cab dataset, we observe a significant difference in speed-up.
\subsection{Ablation Study}
In this experiment, we observe how the building blocks of SLIM work together and provide robustness against changing spatio-temporal level of detail. This includes the effect of using mutually nearest and furthest neighbors, normalization, and IDF components via varying the spatial detail and window width. Figures~\ref{fig:ablation_spatial}~and~\ref{fig:ablation_temporal} show the $F1$-Score as a function of the spatial detail and window width, respectively. Each line represents a different modification to SLIM. 

SLIM matches mutually nearest neighbor event pairs in the similarity computation. It also has an optional step in which the mutually furthest pairs are computed to further support the alibi detection. To compare this approach with other blocking techniques we use two lines. First, the purple line with circle markers (MNN) represents the case where this optional step is removed. Second, the blue line with star markers (All Pairs) represents the case where we match all pairs of events. From Figure~\ref{fig:ablation_spatial}, we observe all three blocking techniques used have similar $F1$-Score values. This is because, the temporal window is narrow ($15$ minutes), and the number of events in each window is already low. At one extreme, if two mobility records have only one event at the same temporal bin, all techniques will return the same pair. Supporting this observation, in Figure~\ref{fig:ablation_temporal} we observe the $F1$-Score of All Pairs decrease dramatically when wider temporal windows are used. For temporal level of $720$ minutes, while the original algorithm and MNN have $0.90$ $F1$-Score, that of All Pairs is only $0.61$. While the effect of the optional MFN step is not obvious for this setting, when we look at the linked pairs, we observe that this step decreases the similarity score of false positive pairs. For spatial level $12$ and temporal level $5$ minutes, the mean of the similarity score of the false positive pairs decreases from $2227$ to $1501$. This indicates that capturing the alibi pairs is harder with MNN only, and the effects of the optional MFN step is significant.

Next, we observe the effects of the normalization and IDF components of the similarity score computation. Red line with cross markers (No IDF) represents the case where IDF is removed, and green line with triangle markers (No Normalization) represents the case where normalization is removed. Figure~\ref{fig:ablation_spatial} shows that the effect of normalization gets more significant with increasing spatial level of detail. 
For spatial level $24$ and temporal level $15$ minutes, we observe $F1$-Score of the original algorithm is $0.96$, while that of No Normalization is $0.76$. When the temporal window is wider, one can observe the significance of the IDF component. For window width $720$ minutes, $F1$-Score of SLIM is $0.89$, while that of No IDF  is $0.69$. This is because, awarding uniqueness becomes even more important, since the unique time-location bins are rarer in wider temporal windows. 

\subsection{Comparison with Existing Work}
\label{sec:exp3}
\noindent We compare SLIM with the two existing approaches. First is GM\footnote{We used the code from the authors: \url{https://tinyurl.com/yagfaz5n}}, which works by learning mobility models from entity records using Gaussian Mixtures and Markov-based models~\cite{ref:gm}. These models are later used for estimating the missing locations of users, and also setting weights to spatio-temporally close record pairs. While we only check records those are in the same temporal window, 
they also award pair of records those are from different temporal windows. 

Second is ST-Link, which performs a sliding-window based comparison over the records of entities and links them if they have $k$ co-occurring records in $l$ diverse locations, and no \emph{alibi} record pairs~\cite{ref:stlink}. 
If an entity has such co-occurring record pairs with more than one entities from the other dataset, all pairs are considered ambiguous and ignored. To identify the values of $k$ and $l$, a trade-off point is identified based on the distribution of all $k$ and $l$-values~\cite{ref:stlink}.

We do not compare our approach with other existing approaches because in its paper GM outperforms eight other approaches, excluding ST-Link. For $F1$-Score comparisons we also include no-LSH SLIM algorithm. For other data points, we apply LSH with $4096$ buckets and $t$ equals $0.6$.

Hit Precision $@k$ is calculated independently for all entities via the formula $1 - max((rank/k), 1)$ and averaged. The $rank$ is the order of the true link in the list of all entities from the opposite dataset, sorted in decreasing order of their similarity score.
Figures~\ref{fig:3-all-hit}~and~\ref{fig:3-all-f1} show the Hit Precision $@40$, running times, and $F1$-Score as a function of the average number of records. Figure~\ref{fig:3-all-f1} also shows the no-LSH SLIM algorithm, represented with single hatched bar and solid line. Since GM does not implement any mechanism to scale to a large number of records, to include it in our results, we took a $1$ week subset of the data. The pivot dataset has $265$ taxis with $675$ records on average. We sampled 5 other datasets, with changing number of average number of records, ranging from $20$ to $660$. These datasets have $265$ taxis, $133$ of which are common with the pivot. With this setting the best achievable hit precision is $0.5$.

From these two figures, we observe that the hit precision values for GM is increasing as the average number of records increases. ST-Link reaches the maximum hit precision with as small as $20$ records. SLIM outperforms GM in all data points, and reaches its best hit precision when the average number of records is $165$. While all three algorithms are able to provide perfect hit precision $@40$, their performance in terms of $F1$-Score differs dramatically. When the average number of records is $20$, SLIM reaches an $F1$-Score of $0.3$, while the other two alternatives stay around $0.05$. Since GM does not link entities with a single entity from the opposite dataset, we apply our linkage and stop threshold algorithm over their similarity scores. While ST-Link is able to rank true positive pairs at the top (we could conclude this from perfect hit precision), it is not always able to detect correct $k$ and $l$ values and resolve ambiguity. When the number of average records per entity increases to $660$, we observe that SLIM still performs the best in terms of the $F1$-Score with $0.92$, while SLIM with LSH shows a similar performance of accuracy with $0.89$ $F1-$ Score. For the same data point, ST-Link and GM have $F1$-Scores of $0.87$ and $0.73$, respectively.

From the running time experiments we observe that GM is two orders of magnitude slower than the other two algorithms. Therefore, we exclude GM in further experiments. Likewise, we exclude no-LSH SLIM algorithm. %We keep ST-Link, as it employs its own scaling techniques. 
Figures~\ref{fig:3-stlink-f1}~and~\ref{fig:3-stlink-speedup} show $F1$-Score, the running times and number of pairwise record comparisons for different record densities, respectively. Green bars correspond to SLIM, the red bars to ST-Link. We use two different intersection amount ratios for each data point. Bars with single hatches show the results for intersection ratio $0.3$, and double hatches show that of $0.7$.
We observe that SLIM outperforms ST-Link in terms of $F1$-Score in all data points except one, and accuracy of ST-Link decreases when the average number of records per entity increases. This is because SLIM is more robust to alibi record pairs than ST-Link, even when alibi threshold mechanism of ST-Link is used. In this experiment, we set the alibi threshold count to $3$ for ST-Link. 

Figure~\ref{fig:3-stlink-speedup} shows that SLIM makes three orders of magnitude less pairwise record comparisons than ST-Link. Likewise, we observe SLIM runs much faster than ST-Link. For average number of records $18,900$, with intersection ration $0.3$, the running time of the SLIM is $343$ seconds, while that of ST-Link takes $1096$ seconds. This is mainly due to the effectiveness of the proposed LSH-based scalability technique. %Also, ST-Link does not scale well in spatially dense areas.
\section{Related Work}
\noindent Many of the attempts in user identity linkage (aka user reconciliation as surveyed in~\cite{ref:linkageSurvey}) utilize additional information such as the network~\cite{ref:korula, ref:alias, ref:molinaICDE15}, profile information (such as usernames or photos)~\cite{ref:gogakdd, ref:uLink,ref:linksocial}, semantic information related to locations~\cite{ref:semantic}, or a combination of these~\cite{ref:hydra,ref:embed}. However, in many cases, only the spatio-temporal information is present or can be used in mobility data, and many of the other identifiers are likely to be anonymized. Use of only spatio-temporal information also aligns better with the new regulations of minimal data collection with consent.

Defining a location based similarity between entities is a fundamental problem \cite{ref:measureSurvey, ref:traMeasure}. Some express this based on densities of location histories~\cite{ref:www9}, either by matching user histograms~\cite{ref:naini}, or using the frequencies of visits to specific locations during specific times~\cite{ref:rossi}. Statistical learning approaches are also used to relate social media datasets with Call Detail Records~\cite{ref:www2,ref:www3}. However, these algorithms depend on discriminative patterns of entities, which is not likely to be present in many datasets, such as the taxi datasets.% we use in this work. 

There have been recent studies to define the similarity among entities using the co-occurrences of their records~\cite{ref:stlink, ref:automatic, ref:kondor, ref:www, ref:gm}. These are the most related group of work to ours, and two of them (ST-Link~\cite{ref:stlink}, GM~\cite{ref:gm}) are included in our experimental evaluation. \emph{SLIM} is shown to outperform them in terms of both accuracy and scalability. Some of these algorithms (GM and Pois~\cite{ref:www}) depend on an assumed model for the users. For example, Pois assumes that visits of each user to a location during a time period follows Poisson distribution and records on each service are independent from each other following Bernoulli distribution. In contrast, we do not make any mobility model assumptions and can work solely on raw spatio-temporal information.
Moreover, these work do not address scalability. We introduce an LSH based solution to scale the linkage process. 

Cao et al.~\cite{ref:automatic} define the strength of the co-occurrences inversely proportional with the frequency of locations. A multi-resolution filtering step is developed for scalability. Different from our approach, the data is pre-processed to add semantic information to locations. 
They do not define a concept of dissimilarity, which is shown to improve both accuracy and efficiency in our solution. Also, they do not automatically determine a similarity threshold to stop linkage.

Trajectory similarity is usually measured using subsequence similarity measures such as the length of the longest common subsequence, Frechet distance, or dynamic time warping\cite{ref:stdm}. 
There are trajectory specific techniques that include information like speed, acceleration, and direction of movement~\cite{ref:jansen}. In contrast to our work, these approaches have strong assumptions, they fell short in addressing asynchrony of the datasets and capturing alibi event pairs. Our approach is more generic as it depends on less features when computing similarity and linkage.
\section{Conclusion}
\noindent In this paper, we studied the problem of identifying the matching entities across mobility datasets using only spatio-temporal information. For this, we first developed a summary representation of mobility records of the entities and a novel way to compute a similarity score among these summaries. This score captures the closeness in time and location of the records, while not penalizing temporal asynchrony. We applied a bipartite matching process to identify the final linked entity pairs, using a stop similarity threshold for the linkage. This threshold is determined by fitting a mixture model over similarity scores to minimize the expected $F1$-Score metric. We also addressed the scalability challenge and employed an LSH based approach for mobility histories, which avoids unnecessary pairwise comparisons. To realize effectiveness of the techniques in practice, we implemented an algorithm called SLIM. Our experiments showed that SLIM outperforms two existing approaches in terms of accuracy and scalability. Moreover, LSH brings two to four orders of magnitude speed-up to the linkage in our experimental settings. 
\balance 
\newpage
\bibliographystyle{ACM-Reference-Format}
\bibliography{sample}


\begin{thebibliography}{49}

%%% the Web address.

\ifx \showCODEN    \undefined \def \showCODEN     #1{\unskip}     \fi
\ifx \showDOI      \undefined \def \showDOI       #1{#1}\fi
\ifx \showISBNx    \undefined \def \showISBNx     #1{\unskip}     \fi
\ifx \showISBNxiii \undefined \def \showISBNxiii  #1{\unskip}     \fi
\ifx \showISSN     \undefined \def \showISSN      #1{\unskip}     \fi
\ifx \showLCCN     \undefined \def \showLCCN      #1{\unskip}     \fi
\ifx \shownote     \undefined \def \shownote      #1{#1}          \fi
\ifx \showarticletitle \undefined \def \showarticletitle #1{#1}   \fi
\ifx \showURL      \undefined \def \showURL       {\relax}        \fi
% invisible to TeX
\providecommand\bibfield[2]{#2}
\providecommand\bibinfo[2]{#2}
\providecommand\natexlab[1]{#1}
\providecommand\showeprint[2][]{arXiv:#2}

\bibitem[\protect\citeauthoryear{Atluri, Karpatne, and Kumar}{Atluri
  et~al\mbox{.}}{2018}]%
        {ref:stdm}
\bibfield{author}{\bibinfo{person}{Gowtham Atluri}, \bibinfo{person}{Anuj
  Karpatne}, {and} \bibinfo{person}{Vipin Kumar}.}
  \bibinfo{year}{2018}\natexlab{}.
\newblock \showarticletitle{Spatio-Temporal Data Mining: A Survey of Problems
  and Methods}. In \bibinfo{booktitle}{\emph{ACM Comp. Surv.}}
  \bibinfo{publisher}{Association for Computing Machinery},
  \bibinfo{address}{New York, NY, USA}, \bibinfo{pages}{83:1--83:41}.
\newblock


\bibitem[\protect\citeauthoryear{Bas{\i}k}{Bas{\i}k}{2017}]%
        {ref:phd}
\bibfield{author}{\bibinfo{person}{Fuat Bas{\i}k}.}
  \bibinfo{year}{2017}\natexlab{}.
\newblock \showarticletitle{Scalable linkage across location enhanced
  services}. In \bibinfo{booktitle}{\emph{CEUR Workshop Proceedings}}. CEUR-WS,
  \bibinfo{publisher}{VLDB Endowment}, \bibinfo{address}{Munich, Germany},
  \bibinfo{pages}{1--4}.
\newblock


\bibitem[\protect\citeauthoryear{Bas{\i}k, Gedik, Etemo\u{g}lu, and
  Ferhatosmano\u{g}lu}{Bas{\i}k et~al\mbox{.}}{2018}]%
        {ref:stlink}
\bibfield{author}{\bibinfo{person}{Fuat Bas{\i}k}, \bibinfo{person}{Bu\u{g}ra
  Gedik}, \bibinfo{person}{Cagri Etemo\u{g}lu}, {and} \bibinfo{person}{Hakan
  Ferhatosmano\u{g}lu}.} \bibinfo{year}{2018}\natexlab{}.
\newblock \showarticletitle{Spatio-Temporal Linkage over Location-Enhanced
  Services}.
\newblock \bibinfo{journal}{\emph{IEEE Transactions on Mobile Computing}}
  \bibinfo{volume}{17}, \bibinfo{number}{2} (\bibinfo{date}{Feb}
  \bibinfo{year}{2018}), \bibinfo{pages}{447--460}.
\newblock


\bibitem[\protect\citeauthoryear{Cao, Wu, Wang, Li, and Wu}{Cao
  et~al\mbox{.}}{2016}]%
        {ref:automatic}
\bibfield{author}{\bibinfo{person}{Wei Cao}, \bibinfo{person}{Zhengwei Wu},
  \bibinfo{person}{Dong Wang}, \bibinfo{person}{Jian. Li}, {and}
  \bibinfo{person}{Haishan Wu}.} \bibinfo{year}{2016}\natexlab{}.
\newblock \showarticletitle{Automatic user identification method across
  heterogeneous mobility data sources}. In \bibinfo{booktitle}{\emph{IEEE Int.
  Conference on Data Engineering (ICDE)}}. \bibinfo{publisher}{IEEE},
  \bibinfo{address}{USA}, \bibinfo{pages}{978--989}.
\newblock


\bibitem[\protect\citeauthoryear{Cecaj, Mamei, and Bicocchi}{Cecaj
  et~al\mbox{.}}{2014}]%
        {ref:www2}
\bibfield{author}{\bibinfo{person}{Alket Cecaj}, \bibinfo{person}{Marco Mamei},
  {and} \bibinfo{person}{Nicola Bicocchi}.} \bibinfo{year}{2014}\natexlab{}.
\newblock \showarticletitle{Re-identification of anonymized CDR datasets using
  social network data}. In \bibinfo{booktitle}{\emph{2014 IEEE International
  Conference on Pervasive Computing and Communication Workshops}}.
  \bibinfo{publisher}{IEEE}, \bibinfo{address}{Budapest, Hungary},
  \bibinfo{pages}{237--242}.
\newblock


\bibitem[\protect\citeauthoryear{Cecaj, Mamei, and Zambonelli}{Cecaj
  et~al\mbox{.}}{2016}]%
        {ref:www3}
\bibfield{author}{\bibinfo{person}{Alket Cecaj}, \bibinfo{person}{Marco Mamei},
  {and} \bibinfo{person}{Franco Zambonelli}.} \bibinfo{year}{2016}\natexlab{}.
\newblock \showarticletitle{Re-identification and information fusion between
  anonymized CDR and social network data}.
\newblock \bibinfo{journal}{\emph{Jour. of Ambient Intelligence and Humanized
  Computing}} \bibinfo{volume}{7}, \bibinfo{number}{1} (\bibinfo{date}{01 Feb}
  \bibinfo{year}{2016}), \bibinfo{pages}{83--96}.
\newblock


\bibitem[\protect\citeauthoryear{{Christen}}{{Christen}}{2012}]%
        {ref:cristenSurvey}
\bibfield{author}{\bibinfo{person}{Peter {Christen}}.}
  \bibinfo{year}{2012}\natexlab{}.
\newblock \showarticletitle{A Survey of Indexing Techniques for Scalable Record
  Linkage and Deduplication}.
\newblock \bibinfo{journal}{\emph{IEEE Transactions on Knowledge and Data
  Engineering}} \bibinfo{volume}{24}, \bibinfo{number}{9} (\bibinfo{date}{Sep.}
  \bibinfo{year}{2012}), \bibinfo{pages}{1537--1555}.
\newblock
\showISSN{2326-3865}


\bibitem[\protect\citeauthoryear{Colizza, Barrat, Barthelemy, and
  Vespignani}{Colizza et~al\mbox{.}}{2006}]%
        {ref:airline}
\bibfield{author}{\bibinfo{person}{Vittoria Colizza}, \bibinfo{person}{Alain
  Barrat}, \bibinfo{person}{Marc Barthelemy}, {and} \bibinfo{person}{Alessandro
  Vespignani}.} \bibinfo{year}{2006}\natexlab{}.
\newblock \showarticletitle{The role of the airline transportation network in
  the prediction and predictability of global epidemics}.
\newblock \bibinfo{journal}{\emph{Proceedings of the National Academy of
  Sciences of the United States of America}} \bibinfo{volume}{103},
  \bibinfo{number}{7} (\bibinfo{year}{2006}), \bibinfo{pages}{2015--2020}.
\newblock


\bibitem[\protect\citeauthoryear{Corless, Gonnet, E.~G.~Hare, Jeffrey, and
  E.~Knuth}{Corless et~al\mbox{.}}{1996}]%
        {ref:LambertW}
\bibfield{author}{\bibinfo{person}{Robert Corless}, \bibinfo{person}{Gaston
  Gonnet}, \bibinfo{person}{D E.~G.~Hare}, \bibinfo{person}{David Jeffrey},
  {and} \bibinfo{person}{D E.~Knuth}.} \bibinfo{year}{1996}\natexlab{}.
\newblock \showarticletitle{On the Lambert W Function}.
\newblock \bibinfo{journal}{\emph{Advances in Computational Mathematics}}
  \bibinfo{volume}{5} (\bibinfo{date}{01} \bibinfo{year}{1996}),
  \bibinfo{pages}{329--359}.
\newblock


\bibitem[\protect\citeauthoryear{de~Montjoye, Hidalgo, Verleysen, and
  Blondel}{de~Montjoye et~al\mbox{.}}{2013}]%
        {ref:4point}
\bibfield{author}{\bibinfo{person}{Yves-Alexandre de Montjoye},
  \bibinfo{person}{C{\'e}sar~A Hidalgo}, \bibinfo{person}{Michel Verleysen},
  {and} \bibinfo{person}{Vincent~D Blondel}.} \bibinfo{year}{2013}\natexlab{}.
\newblock \showarticletitle{Unique in the Crowd: The privacy bounds of human
  mobility}.
\newblock \bibinfo{journal}{\emph{Scientific reports}}  \bibinfo{volume}{3}
  (\bibinfo{year}{2013}), \bibinfo{pages}{1376}.
\newblock


\bibitem[\protect\citeauthoryear{Frias-Martinez, Williamson, and
  Frias-Martinez}{Frias-Martinez et~al\mbox{.}}{2011}]%
        {ref:virus}
\bibfield{author}{\bibinfo{person}{E. Frias-Martinez}, \bibinfo{person}{G.
  Williamson}, {and} \bibinfo{person}{V. Frias-Martinez}.}
  \bibinfo{year}{2011}\natexlab{}.
\newblock \showarticletitle{An Agent-Based Model of Epidemic Spread Using Human
  Mobility and Social Network Information}. In \bibinfo{booktitle}{\emph{IEEE
  Third International Conference on Privacy, Security, Risk and Trust and IEEE
  Third International Conference on Social Computing}}.
  \bibinfo{publisher}{IEEE}, \bibinfo{address}{Boston, Massachusetts, USA},
  \bibinfo{pages}{57--64}.
\newblock


\bibitem[\protect\citeauthoryear{Ganti, Ye, and Lei}{Ganti
  et~al\mbox{.}}{2011}]%
        {ref:crowdsensing}
\bibfield{author}{\bibinfo{person}{Raghu~K. Ganti}, \bibinfo{person}{Fan Ye},
  {and} \bibinfo{person}{Hui Lei}.} \bibinfo{year}{2011}\natexlab{}.
\newblock \showarticletitle{Mobile crowdsensing: current state and future
  challenges}.
\newblock \bibinfo{journal}{\emph{IEEE Communications Magazine}}
  \bibinfo{volume}{49}, \bibinfo{number}{11} (\bibinfo{date}{November}
  \bibinfo{year}{2011}), \bibinfo{pages}{32--39}.
\newblock


\bibitem[\protect\citeauthoryear{Goga, Lei, Parthasarathi, Friedland, Sommer,
  and Teixeira}{Goga et~al\mbox{.}}{2013}]%
        {ref:www9}
\bibfield{author}{\bibinfo{person}{Oana Goga}, \bibinfo{person}{Howard Lei},
  \bibinfo{person}{Sree Hari~Krishnan Parthasarathi}, \bibinfo{person}{Gerald
  Friedland}, \bibinfo{person}{Robin Sommer}, {and} \bibinfo{person}{Renata
  Teixeira}.} \bibinfo{year}{2013}\natexlab{}.
\newblock \showarticletitle{Exploiting innocuous activity for correlating users
  across sites}. In \bibinfo{booktitle}{\emph{Proceedings of the 22nd
  International Conference on World Wide Web}}. \bibinfo{publisher}{Association
  for Computing Machinery}, \bibinfo{address}{New York, NY, USA},
  \bibinfo{pages}{447--458}.
\newblock


\bibitem[\protect\citeauthoryear{Goga, Loiseau, Sommer, Teixeira, and
  Gummadi}{Goga et~al\mbox{.}}{2015}]%
        {ref:gogakdd}
\bibfield{author}{\bibinfo{person}{Oana Goga}, \bibinfo{person}{Patrick
  Loiseau}, \bibinfo{person}{Robin Sommer}, \bibinfo{person}{Renata Teixeira},
  {and} \bibinfo{person}{Krishna~P. Gummadi}.} \bibinfo{year}{2015}\natexlab{}.
\newblock \showarticletitle{On the Reliability of Profile Matching Across Large
  Online Social Networks}. In \bibinfo{booktitle}{\emph{International
  Conference on Knowledge Discovery and Data Mining (KDD)}}.
  \bibinfo{publisher}{Association for Computing Machinery},
  \bibinfo{address}{New York, NY, USA}, \bibinfo{pages}{1799--1808}.
\newblock


\bibitem[\protect\citeauthoryear{Hristova, Williams, Musolesi, Panzarasa, and
  Mascolo}{Hristova et~al\mbox{.}}{2016}]%
        {ref:musolesiwww16}
\bibfield{author}{\bibinfo{person}{Desislava Hristova},
  \bibinfo{person}{Matthew~J. Williams}, \bibinfo{person}{Mirco Musolesi},
  \bibinfo{person}{Pietro Panzarasa}, {and} \bibinfo{person}{Cecilia Mascolo}.}
  \bibinfo{year}{2016}\natexlab{}.
\newblock \showarticletitle{Measuring Urban Social Diversity Using
  Interconnected Geo-Social Networks}. In \bibinfo{booktitle}{\emph{Proceedings
  of the 25th Int. Conference on World Wide Web}}.
  \bibinfo{publisher}{Association for Computing Machinery},
  \bibinfo{address}{New York, NY, USA}, \bibinfo{pages}{21--30}.
\newblock


\bibitem[\protect\citeauthoryear{Indyk and Motwani}{Indyk and Motwani}{1998}]%
        {ref:lsh}
\bibfield{author}{\bibinfo{person}{Piotr Indyk} {and} \bibinfo{person}{Rajeev
  Motwani}.} \bibinfo{year}{1998}\natexlab{}.
\newblock \showarticletitle{Approximate Nearest Neighbors: Towards Removing the
  Curse of Dimensionality}. In \bibinfo{booktitle}{\emph{Proceedings of the
  Thirtieth Annual ACM Symposium on Theory of Computing}}
  \emph{(\bibinfo{series}{STOC '98})}. \bibinfo{publisher}{ACM},
  \bibinfo{address}{New York, NY, USA}, \bibinfo{pages}{604--613}.
\newblock


\bibitem[\protect\citeauthoryear{Kanza, Kravi, Safra, and Sagiv}{Kanza
  et~al\mbox{.}}{2017}]%
        {ref:measureSurvey}
\bibfield{author}{\bibinfo{person}{Yaron Kanza}, \bibinfo{person}{Elad Kravi},
  \bibinfo{person}{Eliyahu Safra}, {and} \bibinfo{person}{Yehoshua Sagiv}.}
  \bibinfo{year}{2017}\natexlab{}.
\newblock \showarticletitle{Location-Based Distance Measures for Geosocial
  Similarity}.
\newblock \bibinfo{journal}{\emph{ACM Transactions on Web}}
  \bibinfo{volume}{11}, \bibinfo{number}{3} (\bibinfo{year}{2017}),
  \bibinfo{pages}{17:1--17:32}.
\newblock


\bibitem[\protect\citeauthoryear{Kieu, Yang, Guo, and Jensen}{Kieu
  et~al\mbox{.}}{2018}]%
        {ref:jansen}
\bibfield{author}{\bibinfo{person}{Tung Kieu}, \bibinfo{person}{Bin Yang},
  \bibinfo{person}{Chenjuan Guo}, {and} \bibinfo{person}{Christian~S. Jensen}.}
  \bibinfo{year}{2018}\natexlab{}.
\newblock \showarticletitle{Distinguishing Trajectories from Different Drivers
  Using Incompletely Labeled Trajectories}. In
  \bibinfo{booktitle}{\emph{Proceedings of the 27th ACM International
  Conference on Information and Knowledge Management}}
  \emph{(\bibinfo{series}{CIKM '18})}. \bibinfo{publisher}{Association for
  Computing Machinery}, \bibinfo{address}{New York, NY, USA},
  \bibinfo{pages}{863--872}.
\newblock


\bibitem[\protect\citeauthoryear{{Kondor}, {Hashemian}, {de Montjoye}, and
  {Ratti}}{{Kondor} et~al\mbox{.}}{2018}]%
        {ref:kondor}
\bibfield{author}{\bibinfo{person}{Daniel {Kondor}}, \bibinfo{person}{Behrooz
  {Hashemian}}, \bibinfo{person}{Yves-Alexandr {de Montjoye}}, {and}
  \bibinfo{person}{Carlo {Ratti}}.} \bibinfo{year}{2018}\natexlab{}.
\newblock \showarticletitle{Towards matching user mobility traces in
  large-scale datasets}.
\newblock \bibinfo{journal}{\emph{IEEE Transactions on Big Data}}
  (\bibinfo{year}{2018}), \bibinfo{pages}{1--1}.
\newblock


\bibitem[\protect\citeauthoryear{Korula and Lattanzi}{Korula and
  Lattanzi}{2014}]%
        {ref:korula}
\bibfield{author}{\bibinfo{person}{Nitish Korula} {and} \bibinfo{person}{Silvio
  Lattanzi}.} \bibinfo{year}{2014}\natexlab{}.
\newblock \showarticletitle{An efficient reconciliation algorithm for social
  networks}.
\newblock \bibinfo{journal}{\emph{VLDB Conference}}  \bibinfo{volume}{7}
  (\bibinfo{year}{2014}), \bibinfo{pages}{377--388}.
\newblock


\bibitem[\protect\citeauthoryear{Kuhn}{Kuhn}{2010}]%
        {ref:hungarian}
\bibfield{author}{\bibinfo{person}{Harold~W. Kuhn}.}
  \bibinfo{year}{2010}\natexlab{}.
\newblock \bibinfo{booktitle}{\emph{The Hungarian Method for the Assignment
  Problem}}.
\newblock \bibinfo{publisher}{Springer Berlin Heidelberg},
  \bibinfo{address}{Berlin, Heidelberg}, \bibinfo{pages}{29--47}.
\newblock


\bibitem[\protect\citeauthoryear{Kurtzberg}{Kurtzberg}{1962}]%
        {ref:approximations}
\bibfield{author}{\bibinfo{person}{Jerome~M. Kurtzberg}.}
  \bibinfo{year}{1962}\natexlab{}.
\newblock \showarticletitle{On Approximation Methods for the Assignment
  Problem}.
\newblock \bibinfo{journal}{\emph{J. ACM}} \bibinfo{volume}{9},
  \bibinfo{number}{4} (\bibinfo{date}{Oct.} \bibinfo{year}{1962}),
  \bibinfo{pages}{419--439}.
\newblock
\showISSN{0004-5411}


\bibitem[\protect\citeauthoryear{Liu, Zhang, Song, Song, Lin, and Hon}{Liu
  et~al\mbox{.}}{2013}]%
        {ref:alias}
\bibfield{author}{\bibinfo{person}{Jing Liu}, \bibinfo{person}{Fan Zhang},
  \bibinfo{person}{Xinying Song}, \bibinfo{person}{Young-In Song},
  \bibinfo{person}{Chin-Yew Lin}, {and} \bibinfo{person}{Hsiao-Wuen Hon}.}
  \bibinfo{year}{2013}\natexlab{}.
\newblock \showarticletitle{What's in a name?: an unsupervised approach to link
  users across communities}. In \bibinfo{booktitle}{\emph{Proceedings of the
  sixth ACM international conference on Web search and data mining}}. ACM,
  \bibinfo{publisher}{Association for Computing Machinery},
  \bibinfo{address}{New York, NY, USA}, \bibinfo{pages}{495--504}.
\newblock


\bibitem[\protect\citeauthoryear{Liu, Wang, and Zhu}{Liu et~al\mbox{.}}{2015}]%
        {ref:semantic}
\bibfield{author}{\bibinfo{person}{Siyuan Liu}, \bibinfo{person}{Shuhui Wang},
  {and} \bibinfo{person}{Feida Zhu}.} \bibinfo{year}{2015}\natexlab{}.
\newblock \showarticletitle{Structured Learning from Heterogeneous Behavior for
  Social Identity Linkage}.
\newblock \bibinfo{journal}{\emph{IEEE Transactions on Knowledge and Data
  Engineering (TKDE)}} \bibinfo{volume}{27}, \bibinfo{number}{7}
  (\bibinfo{year}{2015}), \bibinfo{pages}{2005--2019}.
\newblock


\bibitem[\protect\citeauthoryear{Liu, Wang, Zhu, Zhang, and Krishnan}{Liu
  et~al\mbox{.}}{2014}]%
        {ref:hydra}
\bibfield{author}{\bibinfo{person}{Siyuan Liu}, \bibinfo{person}{Shuhui Wang},
  \bibinfo{person}{Feida Zhu}, \bibinfo{person}{Jinbo Zhang}, {and}
  \bibinfo{person}{Ramayya Krishnan}.} \bibinfo{year}{2014}\natexlab{}.
\newblock \showarticletitle{HYDRA: large-scale social identity linkage via
  heterogeneous behavior modeling}. In \bibinfo{booktitle}{\emph{Proceedings of
  the 2014 ACM SIGMOD International Conference on Management of Data}}.
  \bibinfo{publisher}{Association for Computing Machinery},
  \bibinfo{address}{New York, NY, USA}, \bibinfo{pages}{51--62}.
\newblock


\bibitem[\protect\citeauthoryear{Mu, Zhu, Lim, Xiao, Wang, and Zhou}{Mu
  et~al\mbox{.}}{2016}]%
        {ref:uLink}
\bibfield{author}{\bibinfo{person}{Xin Mu}, \bibinfo{person}{Feida Zhu},
  \bibinfo{person}{Ee-Peng Lim}, \bibinfo{person}{Jing Xiao},
  \bibinfo{person}{Jianzong Wang}, {and} \bibinfo{person}{Zhi-Hua Zhou}.}
  \bibinfo{year}{2016}\natexlab{}.
\newblock \showarticletitle{User Identity Linkage by Latent User Space
  Modelling}. In \bibinfo{booktitle}{\emph{Proceedings of the 22nd ACM SIGKDD
  International Conference on Knowledge Discovery and Data Mining}}.
  \bibinfo{publisher}{Association for Computing Machinery},
  \bibinfo{address}{New York, NY, USA}, \bibinfo{pages}{1775--1784}.
\newblock


\bibitem[\protect\citeauthoryear{Otsu}{Otsu}{1979}]%
        {ref:otsu}
\bibfield{author}{\bibinfo{person}{Nobuyuki Otsu}.}
  \bibinfo{year}{1979}\natexlab{}.
\newblock \showarticletitle{A Threshold Selection Method from Gray-Level
  Histograms}.
\newblock \bibinfo{journal}{\emph{IEEE Transactions on Systems, Man, and
  Cybernetics}} \bibinfo{volume}{9}, \bibinfo{number}{1}
  (\bibinfo{year}{1979}), \bibinfo{pages}{62--66}.
\newblock


\bibitem[\protect\citeauthoryear{Pentland}{Pentland}{2009}]%
        {ref:pentland2009}
\bibfield{author}{\bibinfo{person}{Alex Pentland}.}
  \bibinfo{year}{2009}\natexlab{}.
\newblock \bibinfo{booktitle}{\emph{Reality Mining of Mobile Communications:
  Toward A New Deal On Data}}.
\newblock \bibinfo{publisher}{Springer US}, \bibinfo{address}{Boston, MA},
  \bibinfo{pages}{1--1}.
\newblock
\showISBNx{978-1-4419-0056-2}


\bibitem[\protect\citeauthoryear{Pentland, Lazer, Brewer, and Heibeck}{Pentland
  et~al\mbox{.}}{2009}]%
        {ref:pentlandHealth}
\bibfield{author}{\bibinfo{person}{Alex Pentland}, \bibinfo{person}{David
  Lazer}, \bibinfo{person}{Devon Brewer}, {and} \bibinfo{person}{Tracy
  Heibeck}.} \bibinfo{year}{2009}\natexlab{}.
\newblock \showarticletitle{Using reality mining to improve public health and
  medicine}.
\newblock \bibinfo{journal}{\emph{Studies in health technology and
  informatics}}  \bibinfo{volume}{149} (\bibinfo{date}{02}
  \bibinfo{year}{2009}), \bibinfo{pages}{93--102}.
\newblock


\bibitem[\protect\citeauthoryear{Rajaraman and Ullman}{Rajaraman and
  Ullman}{2011}]%
        {ref:momd}
\bibfield{author}{\bibinfo{person}{Anand Rajaraman} {and}
  \bibinfo{person}{Jeffrey~David Ullman}.} \bibinfo{year}{2011}\natexlab{}.
\newblock \bibinfo{booktitle}{\emph{Mining of Massive Datasets}}.
\newblock \bibinfo{publisher}{Cambridge University Press},
  \bibinfo{address}{New York, NY, USA}. 73--129 pages.
\newblock
\showISBNx{1107015359, 9781107015357}


\bibitem[\protect\citeauthoryear{Reynolds}{Reynolds}{2009}]%
        {ref:gmm}
\bibfield{author}{\bibinfo{person}{Douglas~A. Reynolds}.}
  \bibinfo{year}{2009}\natexlab{}.
\newblock \showarticletitle{Gaussian Mixture Models}. In
  \bibinfo{booktitle}{\emph{Encyclopedia of Biometric Recognition}}.
  \bibinfo{publisher}{Springer}.
\newblock


\bibitem[\protect\citeauthoryear{Riederer, Kim, Chaintreau, Korula, and
  Lattanzi}{Riederer et~al\mbox{.}}{2016}]%
        {ref:www}
\bibfield{author}{\bibinfo{person}{Christopher Riederer},
  \bibinfo{person}{Yunsung Kim}, \bibinfo{person}{Augustin Chaintreau},
  \bibinfo{person}{Nitish Korula}, {and} \bibinfo{person}{Silvio Lattanzi}.}
  \bibinfo{year}{2016}\natexlab{}.
\newblock \showarticletitle{Linking Users Across Domains with Location Data:
  Theory and Validation}. In \bibinfo{booktitle}{\emph{Proc. of the 25th Int.
  Conf.on World Wide Web}}. \bibinfo{publisher}{Association for Computing
  Machinery}, \bibinfo{address}{New York, NY, USA}, \bibinfo{pages}{707--719}.
\newblock
\showISBNx{978-1-4503-4143-1}


\bibitem[\protect\citeauthoryear{Robertson and Walker}{Robertson and
  Walker}{1994}]%
        {ref:bm25}
\bibfield{author}{\bibinfo{person}{Stephen.~E. Robertson} {and}
  \bibinfo{person}{Steve Walker}.} \bibinfo{year}{1994}\natexlab{}.
\newblock \showarticletitle{Some Simple Effective Approximations to the
  2-Poisson Model for Probabilistic Weighted Retrieval}. In
  \bibinfo{booktitle}{\emph{SIGIR '94}}. \bibinfo{publisher}{Springer London},
  \bibinfo{address}{London}, \bibinfo{pages}{232--241}.
\newblock


\bibitem[\protect\citeauthoryear{Rodrigues, Boukerche, Silva, Loureiro, and
  Villas}{Rodrigues et~al\mbox{.}}{2018}]%
        {ref:combTaxi}
\bibfield{author}{\bibinfo{person}{Diego~O. Rodrigues},
  \bibinfo{person}{Azzedine Boukerche}, \bibinfo{person}{Thiago~H. Silva},
  \bibinfo{person}{Antonio~A.F. Loureiro}, {and} \bibinfo{person}{Leandro~A.
  Villas}.} \bibinfo{year}{2018}\natexlab{}.
\newblock \showarticletitle{Combining taxi and social media data to explore
  urban mobility issues}.
\newblock \bibinfo{journal}{\emph{Computer Communications}}
  \bibinfo{volume}{132} (\bibinfo{year}{2018}), \bibinfo{pages}{111 -- 125}.
\newblock


\bibitem[\protect\citeauthoryear{Rossi and Musolesi}{Rossi and
  Musolesi}{2014}]%
        {ref:rossi}
\bibfield{author}{\bibinfo{person}{Luca Rossi} {and} \bibinfo{person}{Mirco
  Musolesi}.} \bibinfo{year}{2014}\natexlab{}.
\newblock \showarticletitle{It's the Way You Check-in: Identifying Users in
  Location-Based Social Networks}. In \bibinfo{booktitle}{\emph{Proceedings of
  the Second ACM Conference on Online Social Networks}}
  \emph{(\bibinfo{series}{COSN '14})}. \bibinfo{publisher}{Association for
  Computing Machinery}, \bibinfo{address}{New York, NY, USA},
  \bibinfo{pages}{215--226}.
\newblock


\bibitem[\protect\citeauthoryear{{Satopaa}, {Albrecht}, {Irwin}, and
  {Raghavan}}{{Satopaa} et~al\mbox{.}}{2011}]%
        {ref:knee}
\bibfield{author}{\bibinfo{person}{Ville {Satopaa}}, \bibinfo{person}{Jeannie
  {Albrecht}}, \bibinfo{person}{David {Irwin}}, {and} \bibinfo{person}{Barath
  {Raghavan}}.} \bibinfo{year}{2011}\natexlab{}.
\newblock \showarticletitle{Finding a "Kneedle" in a Haystack: Detecting Knee
  Points in System Behavior}. In \bibinfo{booktitle}{\emph{2011 31st
  International Conference on Distributed Computing Systems Workshops}}.
  \bibinfo{publisher}{IEEE}, \bibinfo{address}{USA}, \bibinfo{pages}{166--171}.
\newblock


\bibitem[\protect\citeauthoryear{Sharma and Dyreson}{Sharma and
  Dyreson}{2018}]%
        {ref:linksocial}
\bibfield{author}{\bibinfo{person}{Vishal Sharma} {and} \bibinfo{person}{Curtis
  Dyreson}.} \bibinfo{year}{2018}\natexlab{}.
\newblock \showarticletitle{LINKSOCIAL: Linking User Profiles Across Multiple
  Social Media Platforms}. In \bibinfo{booktitle}{\emph{2018 IEEE International
  Conference on Big Knowledge (ICBK)}}. \bibinfo{publisher}{IEEE},
  \bibinfo{address}{USA}, \bibinfo{pages}{260--267}.
\newblock


\bibitem[\protect\citeauthoryear{Shu, Wang, Tang, Zafarani, and Liu}{Shu
  et~al\mbox{.}}{2016}]%
        {ref:linkageSurvey}
\bibfield{author}{\bibinfo{person}{Kai Shu}, \bibinfo{person}{Suhang Wang},
  \bibinfo{person}{Jiliang Tang}, \bibinfo{person}{Reza Zafarani}, {and}
  \bibinfo{person}{Huan Liu}.} \bibinfo{year}{2016}\natexlab{}.
\newblock \showarticletitle{User Identity Linkage across Online Social
  Networks: A Review}.
\newblock \bibinfo{journal}{\emph{SIGKDD Explorations}}  \bibinfo{volume}{18}
  (\bibinfo{year}{2016}), \bibinfo{pages}{5--17}.
\newblock


\bibitem[\protect\citeauthoryear{Steele, Sunds{\o}y, Pezzulo, Alegana, Bird,
  Blumenstock, Bjelland, Eng{\o}-Monsen, de~Montjoye, Iqbal, Hadiuzzaman, Lu,
  Wetter, Tatem, and Bengtsson}{Steele et~al\mbox{.}}{2017}]%
        {ref:poverty}
\bibfield{author}{\bibinfo{person}{Jessica~E. Steele}, \bibinfo{person}{P{\r
  a}l~Roe Sunds{\o}y}, \bibinfo{person}{Carla Pezzulo},
  \bibinfo{person}{Victor~A. Alegana}, \bibinfo{person}{Tomas~J. Bird},
  \bibinfo{person}{Joshua Blumenstock}, \bibinfo{person}{Johannes Bjelland},
  \bibinfo{person}{Kenth Eng{\o}-Monsen}, \bibinfo{person}{Yves-Alexandre de
  Montjoye}, \bibinfo{person}{Asif~M. Iqbal}, \bibinfo{person}{Khandakar~N.
  Hadiuzzaman}, \bibinfo{person}{Xin Lu}, \bibinfo{person}{Erik Wetter},
  \bibinfo{person}{Andrew~J. Tatem}, {and} \bibinfo{person}{Linus Bengtsson}.}
  \bibinfo{year}{2017}\natexlab{}.
\newblock \showarticletitle{Mapping poverty using mobile phone and satellite
  data}.
\newblock \bibinfo{journal}{\emph{Journal of The Royal Society Interface}}
  \bibinfo{volume}{14}, \bibinfo{number}{127} (\bibinfo{year}{2017}),
  \bibinfo{pages}{20160690}.
\newblock


\bibitem[\protect\citeauthoryear{Steorts, Ventura, Sadinle, and
  Fienberg}{Steorts et~al\mbox{.}}{2014}]%
        {ref:blockingSurvey}
\bibfield{author}{\bibinfo{person}{Rebecca~C. Steorts},
  \bibinfo{person}{Samuel~L. Ventura}, \bibinfo{person}{Mauricio Sadinle},
  {and} \bibinfo{person}{Stephen~E. Fienberg}.}
  \bibinfo{year}{2014}\natexlab{}.
\newblock \showarticletitle{A Comparison of Blocking Methods for Record
  Linkage}. In \bibinfo{booktitle}{\emph{Privacy in Statistical Databases}},
  \bibfield{editor}{\bibinfo{person}{Josep Domingo-Ferrer}} (Ed.).
  \bibinfo{publisher}{Springer International Publishing},
  \bibinfo{address}{Cham}, \bibinfo{pages}{253--268}.
\newblock


\bibitem[\protect\citeauthoryear{Unnikrishnan and Naini}{Unnikrishnan and
  Naini}{2013}]%
        {ref:naini}
\bibfield{author}{\bibinfo{person}{Jayakrishnan Unnikrishnan} {and}
  \bibinfo{person}{Farid~Movahedi Naini}.} \bibinfo{year}{2013}\natexlab{}.
\newblock \showarticletitle{De-anonymizing private data by matching
  statistics}.
\newblock \bibinfo{journal}{\emph{2013 51st Annual Allerton Conference on
  Communication, Control, and Computing (Allerton)}} (\bibinfo{year}{2013}),
  \bibinfo{pages}{1616--1623}.
\newblock


\bibitem[\protect\citeauthoryear{Vesdapunt and Garcia-Molina}{Vesdapunt and
  Garcia-Molina}{2015}]%
        {ref:molinaICDE15}
\bibfield{author}{\bibinfo{person}{Norases Vesdapunt} {and}
  \bibinfo{person}{Hector Garcia-Molina}.} \bibinfo{year}{2015}\natexlab{}.
\newblock \showarticletitle{Identifying users in social networks with limited
  information}. In \bibinfo{booktitle}{\emph{IEEE Int. Conference on Data
  Engineering (ICDE)}}. \bibinfo{publisher}{IEEE}, \bibinfo{address}{USA},
  \bibinfo{pages}{627--638}.
\newblock


\bibitem[\protect\citeauthoryear{Wang, Gao, Li, Wang, Jin, and Sun}{Wang
  et~al\mbox{.}}{2018a}]%
        {ref:gm}
\bibfield{author}{\bibinfo{person}{Huandong Wang}, \bibinfo{person}{Chen Gao},
  \bibinfo{person}{Yong Li}, \bibinfo{person}{Gang Wang},
  \bibinfo{person}{Depeng Jin}, {and} \bibinfo{person}{Jingbo Sun}.}
  \bibinfo{year}{2018}\natexlab{a}.
\newblock \showarticletitle{De-anonymization of Mobility Trajectories:
  Dissecting the Gaps between Theory and Practice}. In
  \bibinfo{booktitle}{\emph{The 25th Annual Network \& Distributed System
  Security Symposium (NDSS)}}. \bibinfo{publisher}{The Internet Society},
  \bibinfo{address}{San Diego, CA, USA}, \bibinfo{pages}{1--15}.
\newblock


\bibitem[\protect\citeauthoryear{Wang, Li, Wang, and Jin}{Wang
  et~al\mbox{.}}{2018b}]%
        {ref:gm2}
\bibfield{author}{\bibinfo{person}{Huandong Wang}, \bibinfo{person}{Yong Li},
  \bibinfo{person}{Gang Wang}, {and} \bibinfo{person}{Depeng Jin}.}
  \bibinfo{year}{2018}\natexlab{b}.
\newblock \showarticletitle{You Are How You Move: Linking Multiple User
  Identities From Massive Mobility Traces}. In
  \bibinfo{booktitle}{\emph{Proceedings of the 2018 SIAM International
  Conference on Data Mining}}. \bibinfo{publisher}{SIAM},
  \bibinfo{pages}{189--197}.
\newblock


\bibitem[\protect\citeauthoryear{Wang, Su, Zheng, Sadiq, and Zhou}{Wang
  et~al\mbox{.}}{2013}]%
        {ref:traMeasure}
\bibfield{author}{\bibinfo{person}{Haozhou Wang}, \bibinfo{person}{Han Su},
  \bibinfo{person}{Kai Zheng}, \bibinfo{person}{Shazia~Wasim Sadiq}, {and}
  \bibinfo{person}{Xiaofang Zhou}.} \bibinfo{year}{2013}\natexlab{}.
\newblock \showarticletitle{An Effectiveness Study on Trajectory Similarity
  Measures}. In \bibinfo{booktitle}{\emph{Australian Database Confrence}}.
  \bibinfo{publisher}{Australian Computer Society, Inc.},
  \bibinfo{address}{AUS}, \bibinfo{pages}{13--22}.
\newblock


\bibitem[\protect\citeauthoryear{Wang, Feng, Chen, Yin, Guo, and Chu}{Wang
  et~al\mbox{.}}{2019}]%
        {ref:embed}
\bibfield{author}{\bibinfo{person}{Yaqing Wang}, \bibinfo{person}{Chunyan
  Feng}, \bibinfo{person}{Ling Chen}, \bibinfo{person}{Hongzhi Yin},
  \bibinfo{person}{Caili Guo}, {and} \bibinfo{person}{Yunfei Chu}.}
  \bibinfo{year}{2019}\natexlab{}.
\newblock \showarticletitle{User identity linkage across social networks via
  linked heterogeneous network embedding}.
\newblock \bibinfo{journal}{\emph{World Wide Web}} \bibinfo{volume}{22},
  \bibinfo{number}{6} (\bibinfo{year}{2019}), \bibinfo{pages}{2611--2632}.
\newblock


\bibitem[\protect\citeauthoryear{Xu, Tu, Li, Zhang, Fu, and Jin}{Xu
  et~al\mbox{.}}{2017}]%
        {ref:trajectoryRecovery}
\bibfield{author}{\bibinfo{person}{Fengli Xu}, \bibinfo{person}{Zhen Tu},
  \bibinfo{person}{Yong Li}, \bibinfo{person}{Pengyu Zhang},
  \bibinfo{person}{Xiaoming Fu}, {and} \bibinfo{person}{Depeng Jin}.}
  \bibinfo{year}{2017}\natexlab{}.
\newblock \showarticletitle{Trajectory Recovery From Ash: User Privacy Is NOT
  Preserved in Aggregated Mobility Data}. In
  \bibinfo{booktitle}{\emph{Proceedings of the 26th International Conference on
  World Wide Web}}. \bibinfo{publisher}{International World Wide Web
  Conferences Steering Committee}, \bibinfo{address}{Geneva, CHE},
  \bibinfo{pages}{1241--1250}.
\newblock
\showISBNx{9781450349130}


\bibitem[\protect\citeauthoryear{Zheng}{Zheng}{2015a}]%
        {ref:mstutorial}
\bibfield{author}{\bibinfo{person}{Yu Zheng}.}
  \bibinfo{year}{2015}\natexlab{a}.
\newblock \showarticletitle{Methodologies for Cross-Domain Data Fusion: An
  Overview}.
\newblock \bibinfo{journal}{\emph{IEEE Transactions on Big Data}}
  \bibinfo{volume}{1}, \bibinfo{number}{1} (\bibinfo{date}{March}
  \bibinfo{year}{2015}), \bibinfo{pages}{16--34}.
\newblock


\bibitem[\protect\citeauthoryear{Zheng}{Zheng}{2015b}]%
        {ref:trajectoryDataMining}
\bibfield{author}{\bibinfo{person}{Yu Zheng}.}
  \bibinfo{year}{2015}\natexlab{b}.
\newblock \showarticletitle{Trajectory Data Mining: An Overview}.
\newblock \bibinfo{journal}{\emph{ACM Trans. Intell. Syst. Technol.}}
  \bibinfo{volume}{6}, \bibinfo{number}{3} (\bibinfo{date}{May}
  \bibinfo{year}{2015}), \bibinfo{pages}{29:1--29:41}.
\newblock


\end{thebibliography}

\end{document}